\documentclass[twocolumn,twocolappendix]{aastex63}
\usepackage[T1]{fontenc}
\usepackage{CJK}
\begin{document}

\begin{CJK*}{UTF8}{gbsn}

\title{Fe \textsc{xii} and Fe \textsc{xiii} Line Widths in the Polar Off-limb Solar Corona up to 1.5 $R_\odot$  }

\author[0000-0003-3908-1330]{Yingjie Zhu (朱英杰)}
\affiliation{Department of Climate and Space Sciences and Engineering, University of Michigan, Ann Arbor, MI 48109, USA}

\author[0000-0002-9465-7470]{Judit Szente}
\affiliation{Department of Climate and Space Sciences and Engineering, University of Michigan, Ann Arbor, MI 48109, USA}

\author[0000-0002-9325-9884]{Enrico Landi}
\affiliation{Department of Climate and Space Sciences and Engineering, University of Michigan, Ann Arbor, MI 48109, USA}

\begin{abstract}
	The nonthermal broadening of spectral lines formed in the solar corona is often used to seek the evidence of Alfv\'en waves propagating in the corona. To have a better understanding of the variation of line widths at different altitudes, we measured the line widths of the strong Fe \textsc{xii} 192.4, 193.5, and 195.1 \mbox{\AA} and Fe \textsc{xiii} 202.0 \mbox{\AA} in an off-limb southern coronal hole up to 1.5 $R_\odot$ observed by the Extreme Ultraviolet Spectrometer (EIS) on board the \textit{Hinode} satellite. We compared our measurements to the predictions from the Alfv\'en Wave Solar Model (AWSoM) and the SPECTRUM module. We found that the Fe \textsc{xii} and Fe \textsc{xiii} line widths first increase monotonically below 1.1 $R_\odot$, and then keep fluctuating between 1.1 and 1.5 $R_\odot$. The synthetic line widths of Fe \textsc{xii} and Fe \textsc{xiii} below 1.3 $R_\odot$ are notably lower than the observed ones. We found that the emission from a streamer in the line of sight significantly contaminates the coronal hole line profiles even up to 1.5 $R_\odot$ both in observations and simulations. We suggest that either the discrepancy between the observations and simulations is caused by insufficient nonthermal broadening at the streamer in the AWSoM simulation or the observations are less affected by the streamer. Our results emphasize the importance of identifying the origin of the coronal EUV emission in off-limb observations.
	 
\end{abstract}
\keywords{Solar coronal waves (1995), Solar coronal lines (2038), Solar coronal heating (1989), Spectroscopy (1558)}

\section{Introduction}
\end{CJK*}

The heating of the million-degree solar corona and the acceleration of supersonic solar winds are two key mysteries that have puzzled solar physicists for decades. A significant number of the theories proposed to answer these two questions are related to two small-scale processes: wave or turbulence dissipation, and magnetic reconnection \citep{Cranmer2019}.  

The direct observation of waves propagating in the solar atmosphere by imaging in the past two decades may provide strong evidence for the wave dissipation theories. Waves have been detected in the transverse displacement of chromospheric spicules \citep{DePontieu2007}, in the off-limb corona \citep{Tomczyk2007}, in the transition region \citep{McIntosh2011}, and in the torsional motions of chromospheric spicules \citep{Srivastava2017}. The nature of the observed waves is still under discussion, however \citep[e.g.,][]{vanDoorsselaere2008, Goossens2009,Goossens2013}.  

The presence of nonthermal broadening in ultraviolet emission lines above the limb has been observed since the 1970s \citep[e.g.,][]{Mariska1978,Mariska1979}.  Alfv\'en waves or other acoustic and magnetohydrodynamics (MHD) waves propagating in the corona are suggested as one of the nonthermal mechanisms that broaden spectral lines \citep{Boland1975,Esser1987}. In particular, \citet{Hassler1990} suggested that nonthermal widths of coronal lines caused by the undamped Alfv\'en wave should increase exponentially with altitude. 

The first measurement of coronal line widths in polar coronal holes as a function of height was made by \citet{Hassler1994}. They found that the Fe \textsc{x} 6374 \mbox{\AA} line width increases monotonically to 1.16 $R_\odot$. Later, measurements of the variation of line widths in coronal holes at distances exceeding 1.16 $R_\odot$ were carried out after the launch of the Solar and Heliospheric Observatory \citep[SOHO;][]{Domingo1995}. Observations made with the Solar Ultraviolet Measurements of Emitted Radiation spectrometer \citep[SUMER;][]{Wilhelm1995} on board SOHO revealed that the line widths first increased with height to about 1.2 $R_\odot$ \citep[e.g.,][]{Banerjee1998}, and then were followed by a plateau \citep[e.g.,][]{Pekunlu2002,Moran2003} or started to decrease \citep[e.g.,][]{O'Shea2003}. The increasing line widths in the lower corona were widely interpreted as the existence of undamped Alfv\'en waves in the lower corona, which carry enough energy to heat the corona and accelerate the solar wind \citep[e.g.,][]{Banerjee1998}. The plateau or the decreasing line widths were also regarded as the evidence of wave dissipation above 1.2 $R_\odot$ \citep[e.g.,][]{OShea2005}. 

However, the SUMER measurements of line widths in the off-limb corona are affected by instrumental stray light. \citet{Dolla2008} suggested that the decrease of the line widths above 1.1 - 1.2 $R_\odot$ can be explained by the effect of the stray light, which hinders any convincing measurements of line widths above 1.2 $R_\odot$ using SUMER. In addition, different line widths are found in other structures of a coronal hole. For example, \citet{Raju2000} found that line widths are narrower in plumes than in interplume regions.

The Extreme Ultraviolet Imaging Spectrometer \citep[EIS;][]{Culhane2007} on board the Hinode satellite \citep{Kosugi2007} was used to measure the extreme ultraviolet (EUV) line widths in the off-limb corona after its launch in 2007. These observations showed increasing line widths below 1.15 $R_\odot$ and confirmed the propagation of undamped waves in lower corona  \citep[e.g.,][]{Banerjee2009}. \citet{Bemporad2012} found that the line widths of Fe \textsc{xii} and \textsc{xiii} lines started to decrease at a distance of $\sim 1.14\ R_\odot$. \citet{Hahn2012} found that the line widths in a polar coronal hole started to decrease between 1.1 and 1.3 solar $R_\odot$ depending on the ion. If the decline of line widths is caused by wave dissipation, it can provide 70\% of the energy required to heat the coronal hole. \citet{Hahn2013} also analyzed the decrease of nonthermal velocity in a polar coronal hole and found that 85\% of the energy carried by the waves dissipates below 1.5 $R_\odot$, which is sufficient to heat the coronal hole. Later observations of a southern coronal hole during eclipse using EIS also confirmed that the line widths tend to decrease at 1.2 $R_\odot$ and that the remaining energy flux beyond 1.4 $R_\odot$ is sufficient to accelerate the solar wind \citep{Hara2019}.

MHD simulations have been applied to investigate the role that Alfv\'en waves play in coronal heating and solar wind acceleration \citep[e.g.,][]{Cranmer2007}. The Alfv\'en Wave Solar Model \citep[AWSoM;][]{vanderHolst2014} is a 3D global model in which the low-frequency Alfv\'en wave turbulence drives the coronal heating and solar wind acceleration. \citet{Oran2017} calculated the line profiles from AWSoM simulations and found that the line widths between $1.04$ and $1.34$ $R_\odot$ show good consistency with SUMER observations, except for the Fe \textsc{xii} 1242 \mbox{\AA} line at higher altitude. A recent study made by \citet{Pant2020} found a non-WKB propagation of transverse waves may produce a nonthermal velocity that first increases with height, then starts to level off, and does not show significant variations.
 
 Despite the wealth of studies dedicated to line width behavior as a function of distance, several problems remain unresolved. First, due to the low signal-to-noise ratio (S/N) in the off-limb spectroscopic observations, previous studies usually adopted a large spatial binning along the slit \citep[e.g.,][32 pixels]{Hahn2012} or were restricted to a limited field of view \citep[FOV; e.g.,][$\sim 1.25\ R_\odot$]{Banerjee1998}. In addition, both the SOHO/SUMER and Hinode/EIS instruments are contaminated by instrument-scattered light, which raises doubts about the line profiles measured far from the solar limb. Furthermore, \citet{Szente2019} modeled the EIS line profiles in the region studied by \citet{Hahn2012} using the AWSoM model and an improved spectral synthesis code.  They found that the synthetic line widths did not show any tendency to decrease between 1.0--1.4 $R_\odot$.
 
 Inspired by these problems, we present the measurements of the Fe \textsc{xii} 192.4 \mbox{\AA}, 193.5 \mbox{\AA}, and 195.1 \mbox{\AA} and Fe \textsc{xiii} 202.0 \mbox{\AA} line widths as a function of height in a southern coronal hole observed by EIS with a 33,600 s long total exposure time. The extremely long exposure time allows us to use a smaller spatial binning and measure the line profiles up to $\sim 1.5\ R_\odot$. We also experiment with different stray-light levels in the line fitting to investigate the stray-light effects in observations. We also compare the results with line profiles modeled by an upgraded version of the AWSoM and SPECTRUM module (van der Holst 2021 et al. in progress). The rest of the paper is organized in the following way: in Section 2 we briefly introduce the methods we used in data reduction and analysis. We present the results from both the observations and simulations in Section 3. We discuss the results and their implications in Section 4. Finally, we summarize this paper in Section 5.

\section{Methodology}
\subsection{Data Reduction}
We investigated the observations of a southern coronal hole made by the EIS on board the Hinode satellite \citep{Kosugi2007} during CR 2107 on March 5, 6, and 11, 2011. The main strength of this dataset is the extremely long exposure time --- over 30,000 s per day. The center of the $2''$ wide $512''$ long slit was pointed at ($0''$, $-1242''$) during the off-limb observations, covering a region from $\sim 1.00\ R_\odot$ to $1.54\ R_\odot$. A few on-disk images are taken when the slit center was pointed at ($0''$, $-842''$) to estimate the stray-light levels in off-limb exposures (see Figure~\ref{fig1}). Finally, we obtained 143 frames of the off-limb spectrum and 9 frames of the on-disk spectrum, each of which has an exposure time of 600 s. The position of the EIS slit during on-disk exposures is shown in Figure~\ref{fig1}.

Part of the dataset was not converted into fits files by the EIS pipeline for technical reasons. A parallel suite of IDL programs that converted individual data packets telemetered down from the satellite into science-ready IDL save data files was developed by the authors. We compared the results of this suite of codes with \texttt{eis\_prep} on the datasets for which both the fits files and the data packets were available. They agree very well. After data reduction, the offset along the y-axis was corrected for by the IDL routine \texttt{eis\_ccd\_offset} \citep{EISNote3}. The 1$\sigma$ error in data numbers (DNs) was determined by Poisson statistics and dark-current (readout) noise \citep{EISNote1}. The slit tilt in each image was then corrected for using the quadratic function described in the EIS software notes \citep{EISNote4}. The radiometric calibration is performed in two steps: first we perform the original laboratory calibration to convert the units from $\mathrm{DN\ s^{-1}}$ into $\mathrm{erg\ s^{-1} \ cm^{-2}\ sr^{-1}\ \mbox{\AA}^{-1}}$, and then the decay of the instrument is corrected for following the calibration work by \citet{Warren2014}. It is important to note that \citet{DelZanna2013} developed an alternative revised intensity calibration correction, which shows some disagreement with the \citet{Warren2014} method we adopted. However, because we are interested in the line widths and the only use we make of the line intensities is to determine the intensity ratios of lines close in wavelength, the choice of intensity calibration correction plays a minor role in the present work. We experimented with many different spatial binnings (see Section~\ref{sec4:spabin} and Figure~\ref{fig10}) and found that by increasing the pixels in each bin, the result did not change, but the noise decreased up to a binning over 16 pixels. Beyond the 16-pixel spatial binning, there was no particular gain in noise, but only loss in resolution. Therefore we binned the data along the slit direction by every 16 pixels to further increase the S/N.

\subsection{Stray-light Correction}
Solar EUV emission from the quiet Sun and active regions can be scattered into the EIS FOV, contaminating the observed line profiles when EIS points above the limb. \citet{ugarte2010} measured a 2\% stray-light level using observations during an eclipse, and this value is widely used in studies of coronal line broadening \citep[e.g.,][]{Hahn2012, Hahn2013}. However, a recent study of an orbital eclipse suggested that the amount of stray light in EIS FOV could be higher than 2\% above the distance of 1.3 $R_\odot$ \citep{Hara2019}. In addition, \citet{Wendeln2018} estimated that about a fraction of 10\% radiation from surrounding active and quiet-Sun regions contaminated an equatorial coronal hole observed on the disk. They suggested that the 2\% stray light value might either be underestimated or dependent on the specific configuration of the observation being analyzed, so that the 2\% value proposed by  \citet{ugarte2010} may not apply to all observations.

Unfortunately, it is difficult to determine the contribution of stray light directly from EIS observations. To estimate the stray-light level, the on-disk portion of the slit in the cross-limb observations was averaged along the slit direction and multiplied by a fraction of 2\%, 4\%, and 10\%. We took these three fractions as possible estimates of the stray light level to estimate the uncertainties introduced by removing the stray light.  

\begin{figure*}
    \centering
    \includegraphics[width=\textwidth]{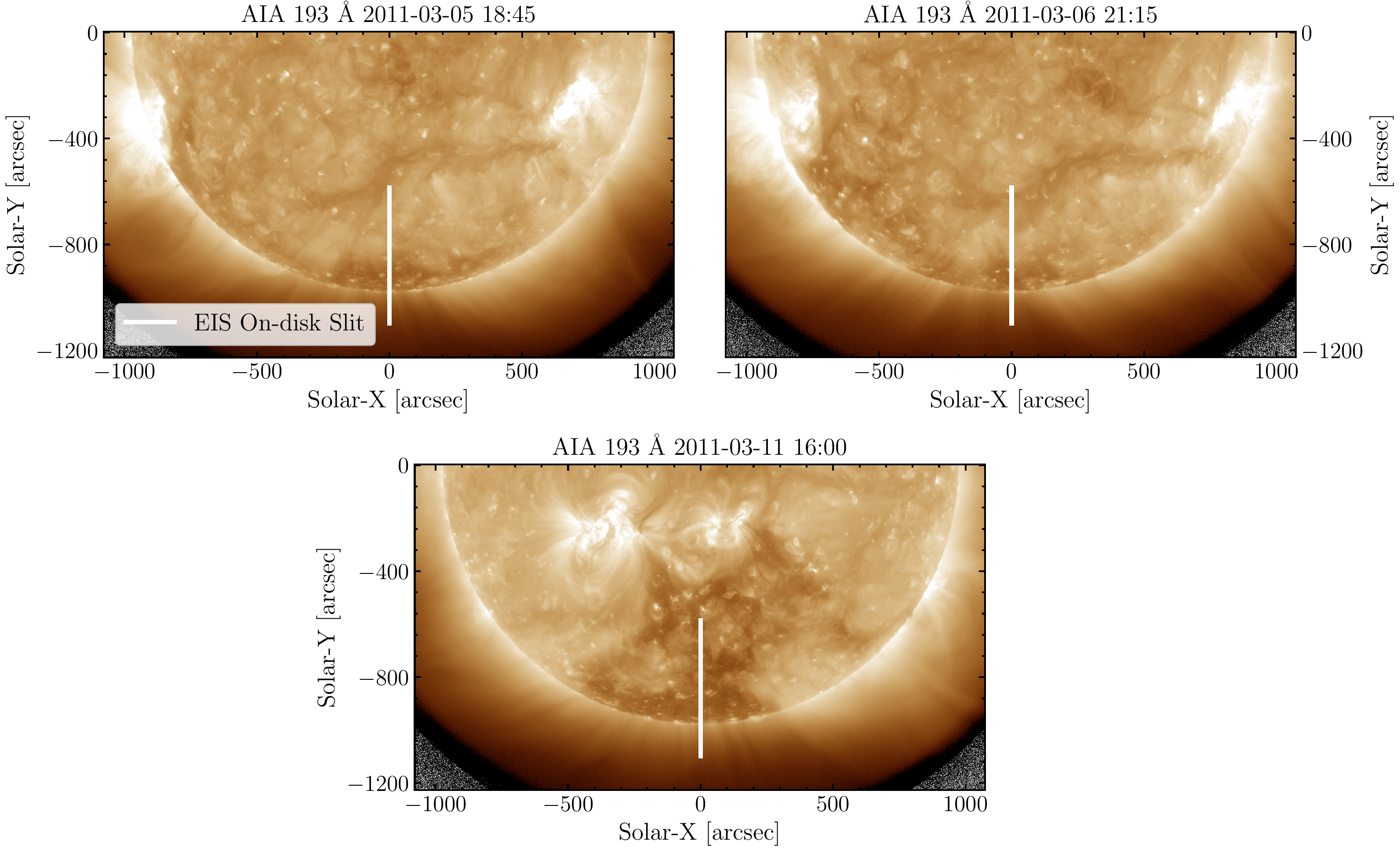}
    \caption{The positions of the EIS Slit over SDO/AIA 193 \mbox{\AA} observations during on-disk exposures on 2011 March 5, 6, and 11. On March 5 and 6, most of the slit was covered by the quiet-Sun plasma. On March 11, the slit was directed at the coronal hole, which explains why the stray-light intensity estimated on March 11 is much lower.}
    \label{fig1}
\end{figure*}

\begin{figure*}
    \centering
    \includegraphics[width=\textwidth]{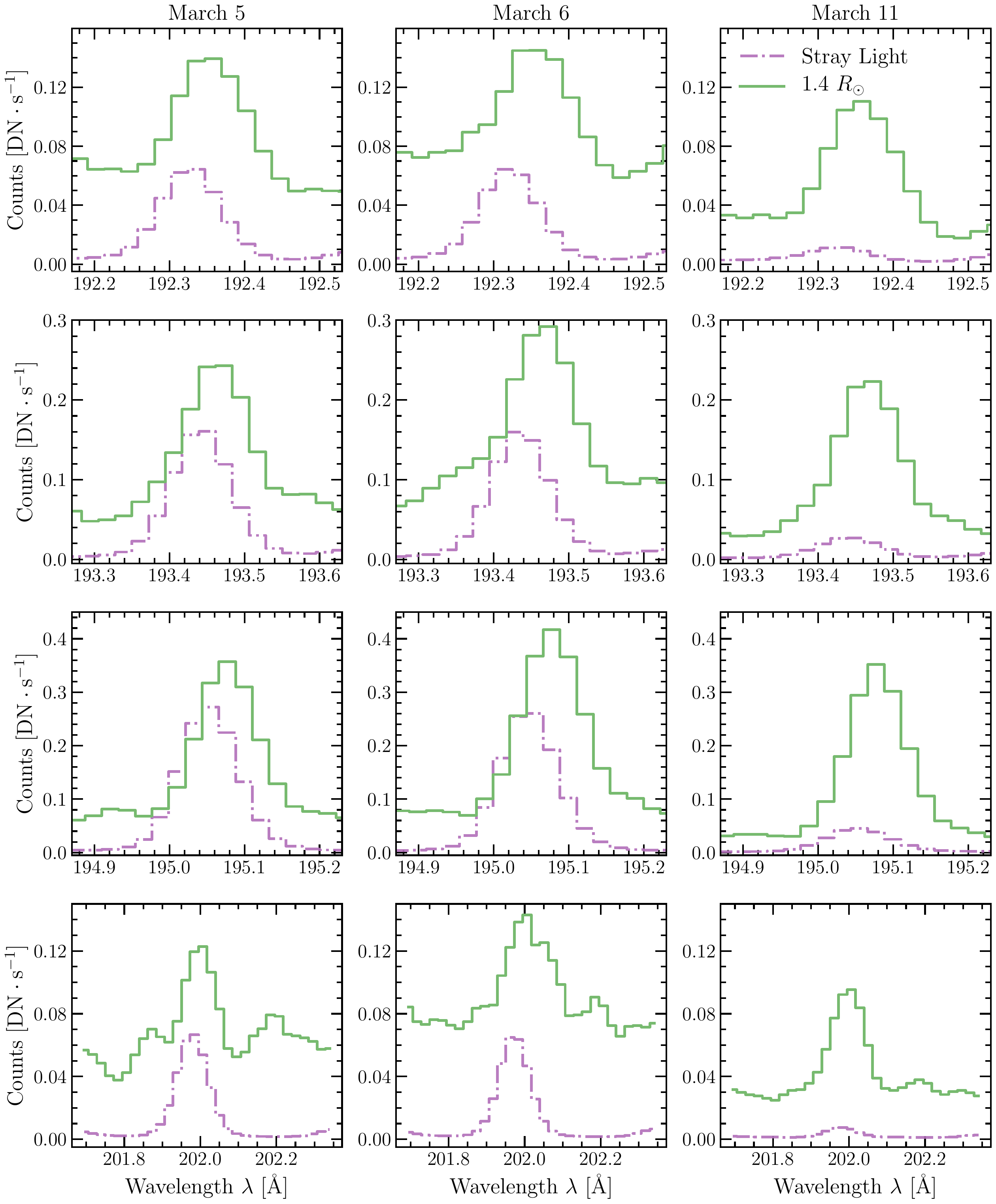}
    \caption{2\% stray-light intensity (dot-dashed) of Fe \textsc{xii} and Fe \textsc{xiii} lines vs. off-limb spectrum (solid) at 1.4 $R_\odot$ on March 5 (left), 6 (middle), and 11 (right). The inferred stray-light contribution on March 11 is much lower than that on March 5 and 6.} 
    \label{fig2}
\end{figure*}

The 2\% stray-light intensity inferred from on-disk exposures on March 5, 6, and 11 are shown in Figure~\ref{fig2} together with the averaged off-limb intensity at$\sim$$1.4\ R_\odot$ on the same date to show the relative strength of the stray light and the measured emission at the slit location where the latter is weakest. The estimated stray light on March 11 contributes much less to the total line intensity than that on the other two days. This is because the slit was pointing to the brighter quiet-Sun plasma during the on-disk exposures on March 5 and 6 (see Figure~\ref{fig1}), rather than the coronal hole on March 11. The stray light was evaluated by fitting a single-Gaussian profile to the on-disk intensity; the fitted profile was rescaled by the stray-light fractions and used to fit the off-limb spectrum, as described in Section~\ref{sec2.3}. We note that there is a small wavelength shift($\sim$$0.01$--$0.02$ \mbox{\AA}) between the line centroids of the on-disk and off-limb spectrum in Figure~\ref{fig2}. We compared the line profiles from the overlapped region ($\sim$1.01--1.12 $R_\odot$) that was covered by on-disk and off-limb exposures and found that the line profiles match perfectly after slit-tilt correction. Therefore we suggest that this is not a systematic shift in the wavelength calibration or slit-tilt correction and use the same line centroid wavelength of the stray-light profiles in the fitting discussed in Section~\ref{sec2.3}. We will further discuss this wavelength shift in Section~\ref{sec4}. 

To avoid the negative DNs when subtracting the estimated stray-light profiles from the off-limb profiles at high altitudes, we discarded the March 5 and 6 data in the rest of the study, leaving 56 off-limb and 3 on-disk exposures. The total exposure time of the off-limb observations is 33,600 s.

\subsection{Fitting}\label{sec2.3}

We fitted observed spectral lines to the summation of two Gaussian profiles with the emcee Markov chain Monte Carlo (MCMC) algorithm \citep{Foreman-Mackey2013}. One component is the "real" off-limb spectrum, and the other is the fixed stray-light profile. The off-limb profile can be determined with four parameters: the integrated intensity $I_0$, the centroid wavelength $\lambda_0$, the full width at half maximum (FWHM) $\Delta \lambda$, and he background intensity $I_{\rm bg}$. 
\begin{eqnarray}
		\ln P(I_{\rm obs}|\lambda, \sigma_I ,I_0, \lambda_0, \Delta \lambda, I_{\rm bg}) =\nonumber \\  -\frac{1}{2} \sum_i \left\{ \left[ \frac{I_{\mathrm{obs}, i} - g(\lambda_i)}{\sigma_i^2} \right]^2 + \ln(2\pi \sigma_i^2)\right\}
\end{eqnarray}
where $g(\lambda_i)$ is the double-Gaussian profile determined by the parameters describing the true emission: $I_0, \lambda_0, \Delta \lambda, I_{\rm bg}$ and the stray-light profile. The $\sigma_i^2$ is determined by the Poisson statistical noise and the CCD readout noise. We used a constant prior in the fitting. The uncertainty of each parameter is defined by $90\%$ credible levels.

\subsection{Instrumental Broadening} 
The FWHM of a Gaussian profile observed by EIS can be written as
\begin{equation}
    \Delta \lambda = \left[\lambda_{\mathrm{I}}^2 + 4\ln 2 \left(\frac{\lambda_0}{c} \right)^2 \left(\frac{2k_B T_i}{m_i} + \xi^2 \right) \right]^{1/2}
\end{equation}
where $\lambda_{\rm I}$ is the instrumental FWHM, $T_i$ is the ion temperature, $m_i$ is the ion mass, $\xi$ denotes the nonthermal widths, $c$ is the speed of light, and $k_B$ is the Boltzmann constant. We calculated the instrumental broadening along the $2''$ slit by using the IDL routine \texttt{eis\_slit\_width.pro} \citep{EISNote7} in the bottom half of the CCD and removed it from the measured values. 


\subsection{AWSoM Simulations and Line Synthesis}
The AWSoM is the representation of the solar corona (SC) and inner heliosphere (IH) components of the Space Weather Modeling Framework \citep[SWMF,][]{Toth2012}. The simulations in this paper were performed with the three-temperature AWSoM model  \citep[AWSoM,][]{vanderHolst2014} where the model solves for isotropic electron and anisotropic proton temperatures. The coronal heating and solar wind acceleration are addressed via low-frequency Alfv\'en waves that are partially reflected by gradients of the solar wind plasma. The energy partitioning of the model has been improved, so that the electron-to-proton heating ratio increases, resulting in decreased, more realistic polar wind speeds (see van der Holst et al. 2021, in progress). This paper uses the SC component, which uses a spherical grid between 1 and 24 $R_\odot$ for which the radial coordinate is stretched to numerically resolve the steep gradients near the Sun. We also artificially broadened the transition region similar to \citet{Lionello2009} and \citet{Sokolov2013}. The domain is decomposed into adaptive mesh refinement (AMR) blocks, and we applied one additional level of AMR around the heliospheric current sheet. To obtain plasma parameters for the times of the observation, we used a Global Oscillation Network Group magnetogram \citep[GONG,][]{Harvey:1996} processed with Air Force Data Assimilative Photospheric flux Transport \citep[][]{Henney:2012} into magnetograms as the radial magnetic field at the inner boundary, while the solar wind initial condition was the Parker solution. Then we obtained a steady-state solution after 100,000 steps, from which we extracted the observed plasma parameters into Cartesian boxes covering the region of emission observed by Hinode/EIS.

The solar wind model is capable of predicting both in-situ and remote-sensing observations. SPECTRUM is a post-processing tool within the SWMF that uses the AWSoM coronal simulation results to calculate synthetic spectra on a voxel-by-voxel basis. To calculate the emergent line profiles, the AWSoM simulation results are first interpolated to a Cartesian grid after the line-of-sight (LOS) direction has been specified by the user. Then the SPECTRUM module performs the calculation voxel by voxel. In each voxel, we calculate the emissivity profiles, which include Doppler shifts due to the plasma motion along the LOS direction (Equation~(\ref{spe:eq:doppler})), and the broadening of the emissivity profiles (Equation~(\ref{spe:linebroadening})). Finally, SPECTRUM integrates the emissivity profiles along the LOS direction to obtain the emergent spectrum.

SPECTRUM takes care of two different Doppler-broadening mechanisms:

\noindent
1) The first is the macroscopic Doppler broadening due to the LOS integration of the line profile with different line centroids. The centroid of the emissivity profile at each voxel is shifted by the local bulk plasma motion $u_{\rm LOS}$
\begin{eqnarray}
  \label{spe:eq:doppler}
  \lambda_{\rm shifted} = \left(1 -\frac{u_{\rm LOS}}{c} \right) \lambda_0 ,
\end{eqnarray}
where $c$ is the speed of light.

\noindent
2) The second mechanism is the microscopic Doppler broadening in individual voxels caused by the thermal motion and Alfv\'enic nonthermal motions,
\begin{eqnarray}
  \label{spe:linebroadening}
      \Delta \lambda = \left[4\ln 2 \left(\frac{\lambda_0}{c} \right)^2 \left(\frac{2 k_B T_{\rm LOS}}{ m_{p} A_{i}}  + \xi^2 \right) \right]^{1/2}
  \end{eqnarray}       
  where $m_p$ is the proton mass, and $A_i$ is the mass number of the ion. Because AWSoM does not yet predict ion temperatures, we assumed the LOS ion temperature to be given by the LOS proton temperature $T_{\rm LOS} = T_{\rm perp} \sin^2 \alpha + T_{\rm par} \cos^2 \alpha$, where $T_{\rm perp}$ and $T_{\rm par}$ are proton temperatures perpendicular and parallel to the magnetic field, and $\alpha$ is the angle between the direction of the local magnetic field and the LOS; this assumption is further discussed in Section~\ref{sec4:iontemp}. 
  
The nonthermal component is 
\begin{eqnarray}
  \label{spe:eq:vnonthermal}
  \xi^2 = \frac{1}{2} \langle \delta u^2 \rangle\sin^2{\alpha} =
  \frac{1}{2}\frac{\omega^{+}+\omega^{-}}{\rho} \sin^2{\alpha} =  \nonumber \\
  \frac{1}{8}  \left(z_{+}^2 + z_{-}^2 \right)\sin^2{\alpha} .  
\end{eqnarray}
Here $z_{\pm}$ are the Els\"{a}sser variables for forward- and
backward-propagating waves, and the respective energy densities are $\omega_{\pm}$.


We note that the nonthermal velocity in this paper is defined in a slightly different but equivalent way in comparison to the definitions in \citet{Szente2019}. \citet{Szente2019} adopted the standard deviation to describe the line broadening, and the nonthermal velocity is defined as $v_{\rm nth}^2 = \frac{1}{4} \langle \delta u^2 \rangle\sin^2{\alpha} = \frac{1}{2} \xi^2$. For the detailed implementation of synthetic spectral calculations, see \citet{Szente2019}.
 
\section{Results}
\begin{figure*}
	\centering
	\gridline{\fig{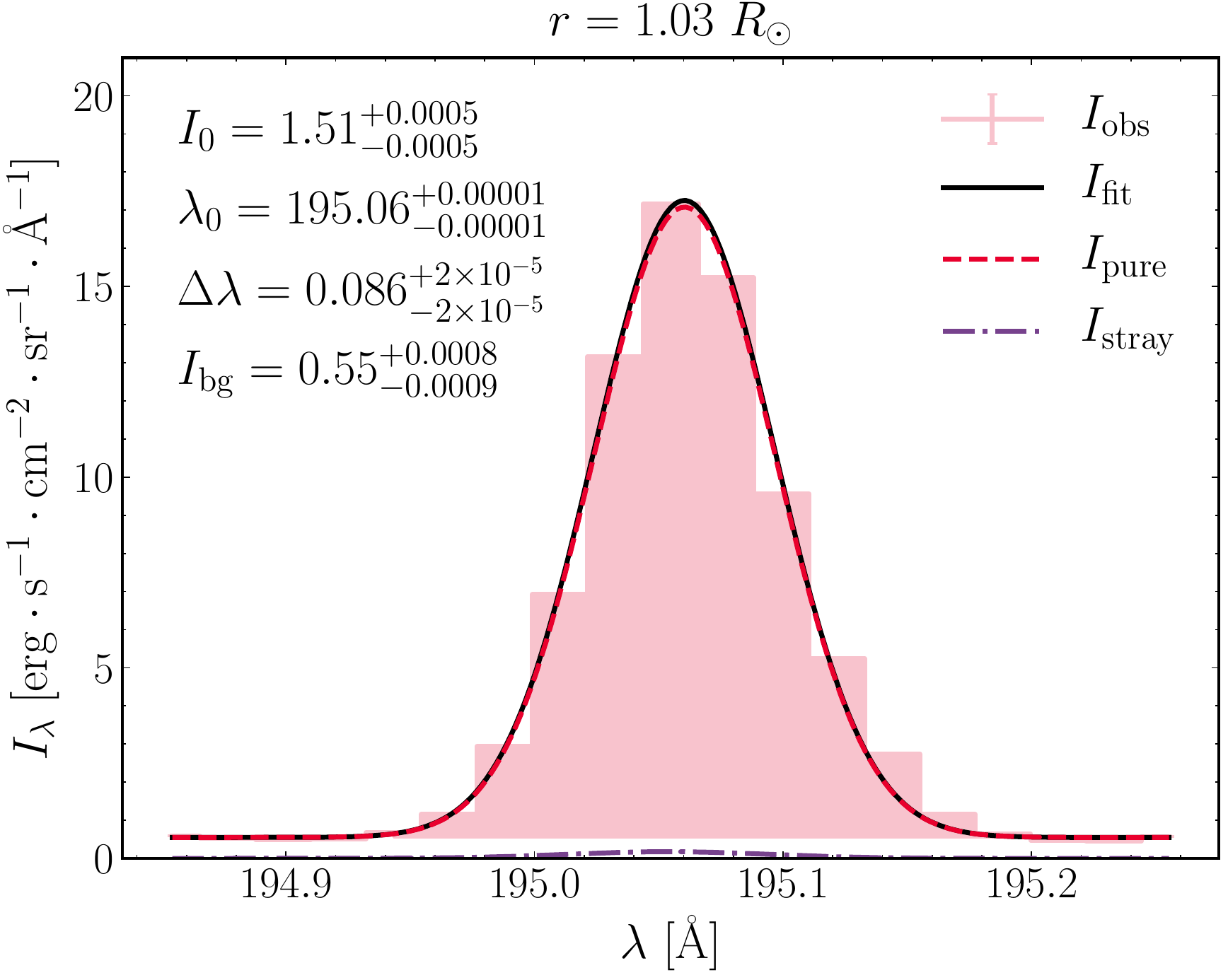}{0.33\textwidth}{(a)}
	\fig{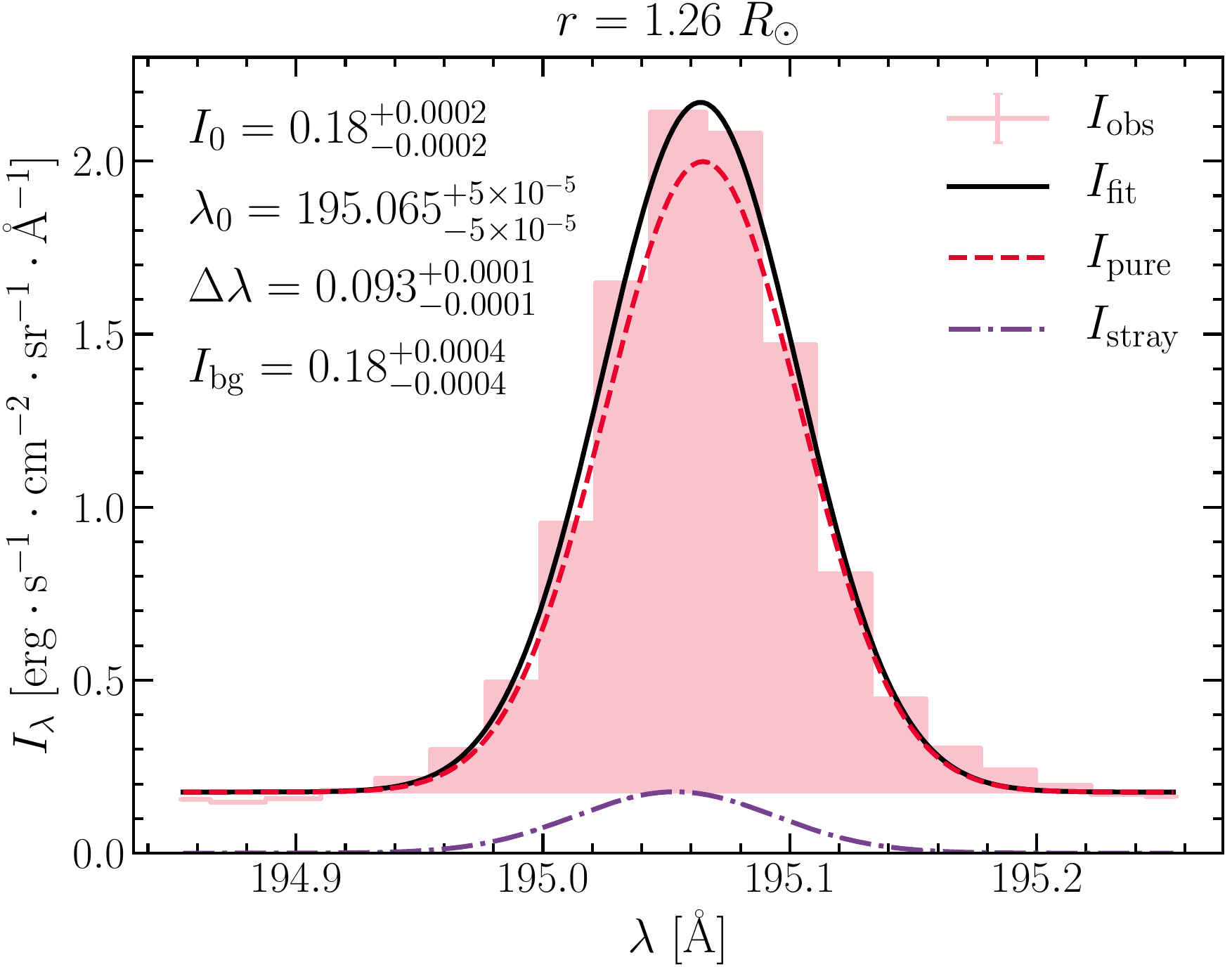}{0.33\textwidth}{(b)}
	\fig{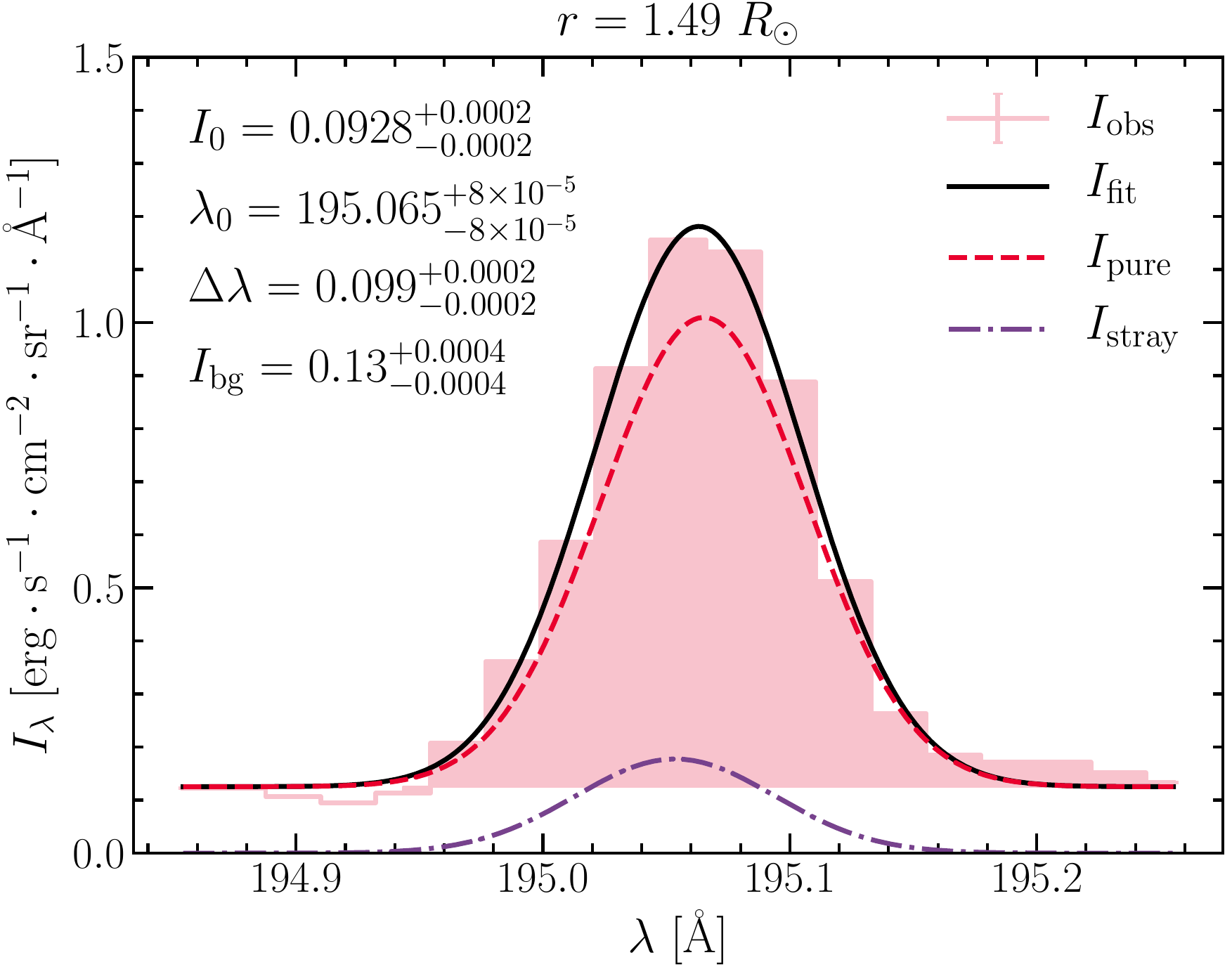}{0.33\textwidth}{(c)}}
	
	\caption{Examples of fitting an observed Fe \textsc{xii} 195.1 \mbox{\AA} line profile at$\sim$$1.03,1.26, \mathrm{and}\ 1.49\ R_\odot$ assuming a stray-light level of 2\%. The pink histogram represents the observed line profiles. The dashed red profiles are the fitted off-limb true emission. The dot-dashed purple lines are the 2\% stray-light profile. The solid black lines are the sum of the fitted off-limb spectrum and the stray-light profile. The fitting parameters and their 90\% credible levels are listed as well (integrated intensity $I_0$, line centroid wavelength $\lambda_0$, FWHM $\Delta \lambda$, and background intensity $I_{\rm bg}$).} 
	\label{fig3} 
\end{figure*}

We first show three examples of fitting a Fe \textsc{xii} 195 \mbox{\AA} line observed at$\sim$$1.03, 1.26, \mathrm{and}\ 1.49\ R_\odot$ assuming a 2\% stray-light level in Figure~\ref{fig3}. The observational uncertainties are too small to be shown in the figure due to the 33,600 s exposure time and the spatial binning. The inferred integrated intensity $I_0$, the line centroid position $\lambda_0$, the FWHM $\Delta \lambda$, and the background intensity $I_{\rm bg}$ are listed in each panel. 
At 1.03 $R_\odot$, the stray-light profile is negligible. At 1.26 $R_\odot$, the stray light only contributes a tiny portion of the intensity at the line core and does not significantly affect the line width. At 1.49 $R_\odot$, the 2\% stray-light intensity is still too low to dominate the off-limb profile. Therefore we can still fit the off-limb spectrum by two Gaussian components with good precision.

\begin{figure}
	\centering
	\includegraphics[width=\linewidth]{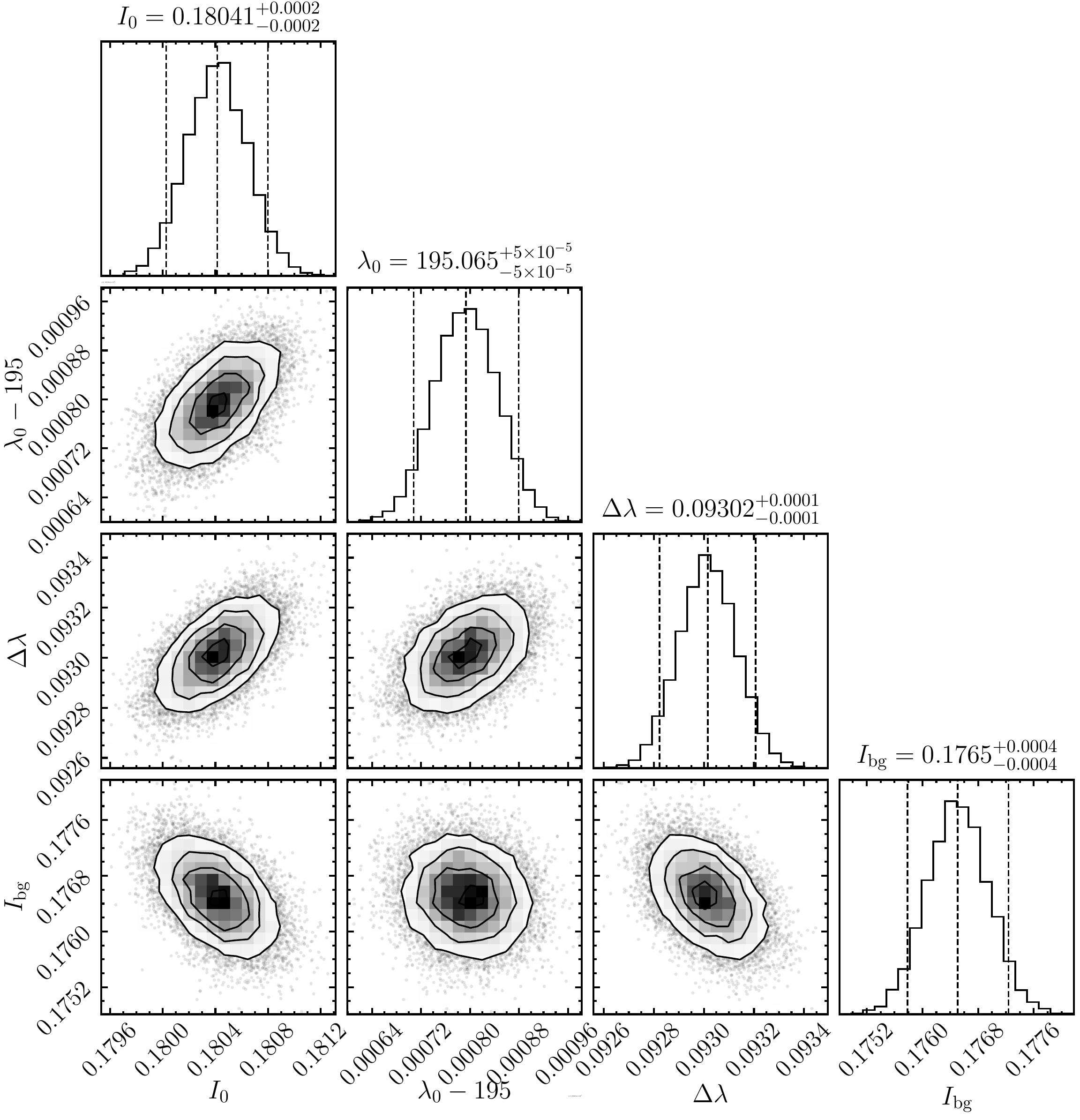}
	\caption{A corner plot of the marginalized distribution of the posterior probability sampled by MCMC algorithms from the Fe \textsc{xii} 195 \mbox{\AA} line profile in Figure~\ref{fig3} (b), corresponding to 1.26 $R_\odot$: $I_0$ integrated intensity, $\lambda_0$ line centroid wavelength, $\Delta \lambda$ FWHM, and $I_{\rm bg}$ background intensity. The diagonal panels show the 1D distribution of the parameters. The vertical lines in each panel stand for the 5\%, 50\%, and 95\% cumulative probability (i.e., 90\% credible levels). The 2D posterior probability distribution between each two parameters is shown in the off-diagonal panels. This figure is generated using the python package \texttt{corner.py} \citep{corner}.}
	\label{fig4}
\end{figure}

The 2D posterior probability distribution of the four parameters, integrated intensity $I_0$, line centroid wavelength $\lambda_0$, FWHM $\Delta \lambda$, and background intensity $I_{\rm bg}$ of the Figure~\ref{fig3}(b) are shown in Figure~\ref{fig4}. The fitted FHWM $\Delta \lambda$ shows some correlations between other parameters. The FWHM $\Delta \lambda$ is positively correlated with the total intensity $I_0$ because the maximum of the Gaussian profile is well determined. Therefore a larger FWHM $\Delta \lambda$ leads to a higher total intensity $I_0$. The correlations between the FHWM $\Delta \lambda$ and the line centroid wavelength $\lambda_0$ implies the underlying asymmetry in the observed line profile. The FWHM negatively correlates with the background intensity $I_{\rm bg}$. This is because the far wings can be fit either with a larger line width $\Delta \lambda$ or with a higher background intensity $I_{\rm bg}$. A higher background level masks the wings and narrows the widths down.


\begin{figure*}
	\centering
	\includegraphics[width=0.6\linewidth]{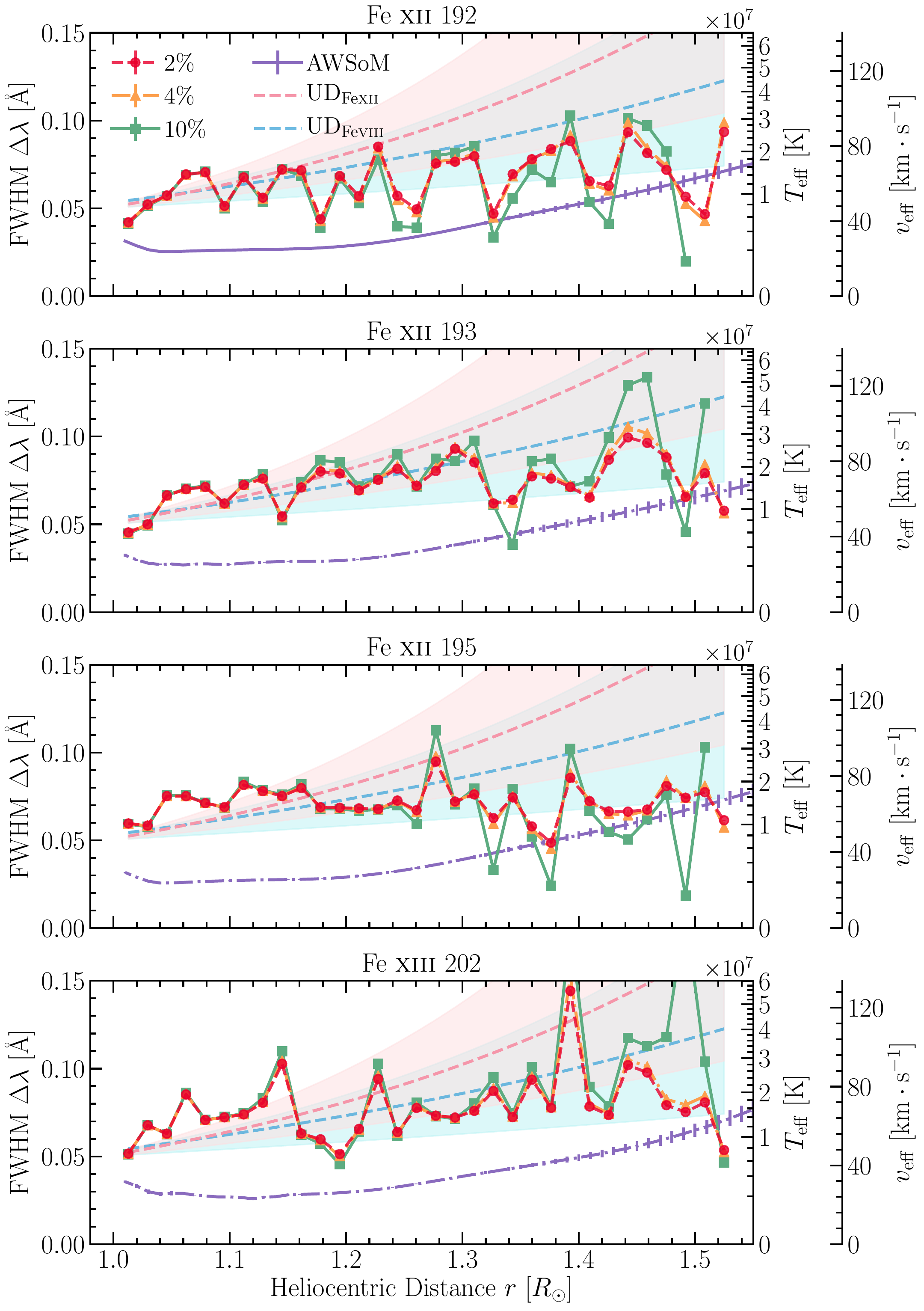}
	\caption{The measured FWHMs and the effective temperature or the effective velocity of the Fe \textsc{xii} 192.4, 193.5, and 195.1 \mbox{\AA} and Fe \textsc{xiii} 202.0 \mbox{\AA} lines as a function of heliocentric distance using different stray-light levels. The data were spatially binned in every 16 pixels along the y-axis. The red, yellow, and green lines illustrate the FWHMs fitted using 2\%, 4\%, and 10\% stray-light levels, respectively. The purple lines represent the line widths inferred from the SPECTRUM synthetic spectrum. The line widths caused by undamped (UD) waves are shown in blue and pink curves, which are inferred from density measurements of Fe \textsc{viii} and Fe \textsc{xii} line pairs, respectively. The undamped curves are normalized at 1.05 $R_\odot$ for the Fe \textsc{xii} 192, 193, and Fe \textsc{xiii} 202 lines. The shaded blue and pink area indicates the uncertainty in the estimation of the undamped widths. Note that the instrumental widths have been subtracted in this figure.}
	\label{fig5}
\end{figure*}

In Figure~\ref{fig5}, we show the measured FWHMs (instrumental widths subtracted) of the Fe \textsc{xii} 192.4, 193.5, and 195.1 \mbox{\AA} and Fe \textsc{xiii} line 202.0 \mbox{\AA} lines as a function of height using different stray-light levels with a 16-pixel spatial binning. The FWHMs estimated from the SPECTRUM synthetic spectrum are also shown as a comparison. The undamped line widths at different heliocentric distance assuming $\Delta \lambda \propto n_e^{-1/4}$ are normalized to each measured FWHM curve at$\sim$$1.05\ R_\odot$ for reference, except for the broader Fe \textsc{xii} 195 \mbox{\AA} line (see the discussion in Section~\ref{sec4:inseff}). We estimated the electron density $n_e$ from two independent line pairs using the CHIANTI database version 9 \citep{Dere1997,Dere2019}: (1) the intensity ratio of Fe \textsc{xii} 195.1 and blended 186.9 \mbox{\AA} lines, and (2) the intensity ratio of Fe \textsc{viii} 185.2 and 186.5 \mbox{\AA} lines. The inferred the electron density below$\sim$$1.1\ R_\odot$ was fit by an exponential function and extrapolated to higher altitudes where the rapidly decreasing Fe \textsc{xii} 186.9 \mbox{\AA} and Fe \textsc{viii} line intensities prevented a reliable measurement of the electron density. The density (pressure) scale heights are $75_{-20}^{+50}\ \mathrm{Mm}$ (Fe \textsc{xii}) and $110_{-35}^{+100}\ \mathrm{Mm}$ (Fe \textsc{viii}), which correspond to scale-height temperatures $T\sim 1.6_{-0.5}^{+1.0}\ \mathrm{MK}$ (Fe \textsc{xii}) and $T\sim 2_{-0.5}^{+2}\ \mathrm{MK}$ (Fe \textsc{viii}) in the hydrostatic case, respectively. The differences in the scale height inferred from the Fe \textsc{viii} and Fe \textsc{xii} line ratios might be due to the different structures from which the photons are emitted.

\begin{figure*}[t]
	\centering
	\includegraphics[width=0.99\textwidth]{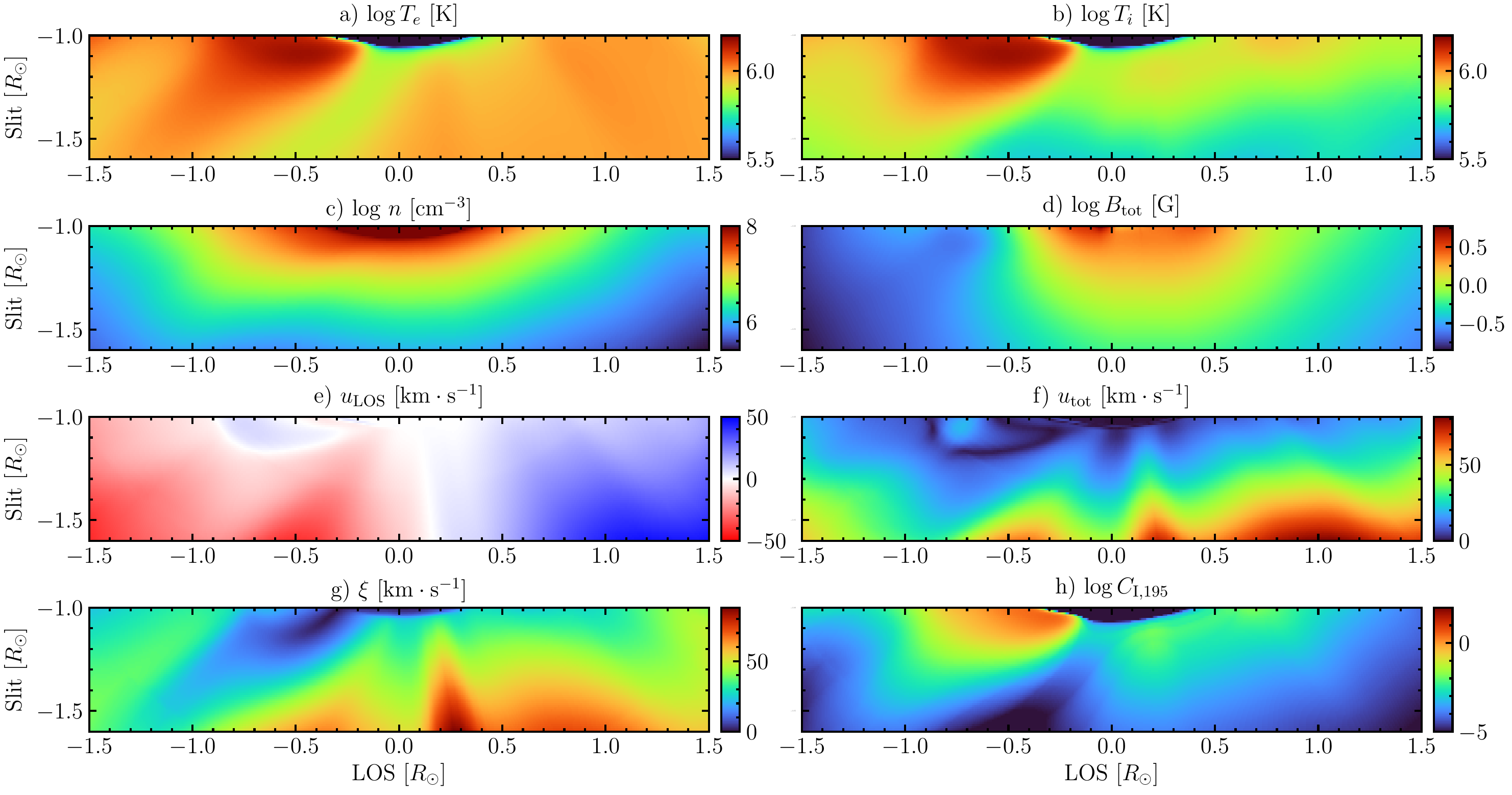}
	\caption{Physical quantities in AWSoM outputs in the meridional cuts of the solar corona taken in the plane corresponding to the LOS, perpendicular to the ecliptic: (a) electron temperature $T_e$; (b) ion temperature $T_i$; (c) particle number density $n$; (d) total magnetic field strength $B_{\rm tot}$; (e) LOS velocity $u_{\rm LOS}$; (f) total velocity $u_{\rm tot}$; (g) local LOS nonthermal velocity $\xi$; and (h) contribution function of the Fe \textsc{xii} 195 \mbox{\AA} line $C_{\rm I, 195}$. }
	\label{fig6}
\end{figure*}

Figure 5 shows several different things. First of all, there is no significant evidence showing that line widths start to decrease between 1.2 $R_\odot$ to 1.5 $R_\odot$ in all the four strong Fe \textsc{xii} and Fe \textsc{xiii} lines. The line widths of Fe \textsc{xii} 192.4 and 193.5 \mbox{\AA} lines first increase from $\sim$$0.04$ \mbox{\AA} to $\sim$$
0.06$ \mbox{\AA} between$\sim$1.0--1.05 $R_\odot$. Then the 192.4 and 193.5 \mbox{\AA} line widths start to fluctuate between 0.05 and 0.10 \mbox{\AA} up to 1.5 $R_\odot$. The Fe \textsc{xii} 195.1 \mbox{\AA} line widths are slightly larger than Fe \textsc{xii} 192.4 and 193.5 \mbox{\AA} line below 1.2 $R_\odot$ by$\sim$0.005--0.01 \mbox{\AA}.  The Fe \textsc{xii} 195.1 \mbox{\AA} line widths also continue to rise and fall between$\sim$0.05--0.1 \mbox{\AA} and we cannot find any systematic decrease in line widths above 1.2 $R_\odot$. The Fe \textsc{xiii} 202.0 \mbox{\AA} line widths vary with height in a similar manner to the Fe \textsc{xii} line widths. At some heights (e.g., $\sim 1.15\ R_\odot, 1.4\ R_\odot$), the Fe \textsc{xiii} 202.0 \mbox{\AA} line is extremely broadened, showing widths larger than 0.1 \mbox{\AA}. The fluctuations in the FWHMs of the four spectral lines have a spatial period of$\sim$0.05--0.1 $R_\odot$ and an amplitude of$\sim$0.01--0.02 \mbox{\AA} ($\sim$10--20 $\mathrm{km\cdot s^{-1}}$ in effective velocity). 
Assuming no nonthermal broadenings, the line width measured in the off-limb spectrum corresponds to an effective temperature of more than $10$ MK. 

Second, different stray-light levels do not affect the fitting results below 1.3 $R_\odot$ because the stray-light contributions to the total intensity are negligible. The line widths inferred from the 2\% and 4\% stray-light level are almost identical at all heights. Above 1.3 $R_\odot$, the line widths measured assuming 10\% stray-light level are significantly different from those from the 2\% or 4\% stray-light level by 0.01--0.05 \mbox{\AA}. When there is no significant wavelength shift between the stray light and the observed profile, a higher stray-light level results in a larger fitted width because the stray-light intensity becomes comparable to or even dominates the total off-limb intensity and maximizes at the line core, so that the core intensity is decreased more than the wing intensity, broadening the line. However, the measured widths in 10\% of the stray-light level can sometimes be smaller than the 2\% level. This is because the wavelength shift between the stray-light and off-limb profile causes the stray-light removal to affect the portion of the observed line profile to the blue wing more than the one at the red wing, causing an artificial narrowing of the line. 

Third, the line widths of the Fe \textsc{xii} and \textsc{xiii} lines measured by EIS are much larger than the synthetic line widths, but are smaller than the normalized undamped widths. The line widths of the AWSoM simulations are $\sim$0.03 \mbox{\AA} below 1.2 $R_\odot$, which is $\sim$0.03--0.04 \mbox{\AA} smaller than the EIS observations. At larger heights, the AWSoM widths begin to increase monotonically with height, from $\sim$0.02 \mbox{\AA} to $\sim$0.07 \mbox{\AA} at 1.6 $R_\odot$. Neither fluctuation nor decrease in line widths is found in the AWSoM results. The undamped widths inferred from Fe \textsc{xii} line ratios grow more rapidly than the Fe \textsc{viii} curves because the measured density scale height is smaller. The undamped line widths, especially the Fe \textsc{viii} curves, are close to the upper limit of the fluctuating EIS widths below 1.4 $R_\odot$ after which they become larger. The AWSoM widths increase at a slightly lower rate than the undamped waves, which implies that the AWSoM may account for some wave dissipation, but less than the observations indicate.

Figure~\ref{fig6} shows the physical quantities of AWSoM simulations in the meridional cuts of the solar corona taken in the plane corresponding to the LOS of the entire EIS slit, perpendicular to the ecliptic. The electron temperature $T_e$ and proton temperature $T_i$ in Figure~\ref{fig6} (a) and (b) have similar distributions. A possible streamer is revealed in the temperature plots with an almost identical electron temperature and ion temperature $\log T \sim 6.1$. The ion temperature $T_i$ in the polar coronal hole ($\log T_i \sim 5.9$) is slightly lower than the electron temperature $T_e$ ($\log T_e \sim 6.0$). The particle density in the coronal hole drops from $\log n \sim 8$ to $\log n \sim 6$ from the limb to 1.5 $R_\odot$. The total magnetic field strength in the polar region is $\sim$1--3 G. The LOS velocity $u_{\rm LOS}$ shown in Figure~\ref{fig6} (e) increases from$\sim$$10\ \mathrm{km \ s^{-1}}$ at the limb to$\sim$$50\ \mathrm{km \ s^{-1}}$ at 1.6 $R_\odot$. The bulk flow in the streamer at the far side of the Sun is moving toward the observer, unlike the other solar wind flows at the far side. The total velocity $u_{\rm tot}$ also increases from$\sim$$20\ \mathrm{km \ s^{-1}}$ to$\sim$$90\ \mathrm{km \ s^{-1}}$ from the limb to 1.6 $R_\odot$. Moreover, neither LOS nor total speed in the coronal hole are symmetrically distributed, but rather show structures in which the wind acceleration is stronger than in the rest of the coronal hole.

In Figure~\ref{fig6} (g), we show the distribution of local nonthermal velocity $\xi$ caused by Alfv\'en waves. 
The distribution of the nonthermal velocity $\xi$ reveals similar fine structures along the LOS, including a plume in the coronal hole where the nonthermal velocity is much smaller. In the streamer, the local nonthermal velocity is only$\sim$$10\ \mathrm{km\ s^{-1}}$ because the local Alfv\'en wave energy density is low. In the polar coronal hole, the nonthermal velocity $\xi$ is much higher, it increases from$\sim 50$$\mathrm{km\ s^{-1}}$ to$\sim$$100\ \mathrm{km\ s^{-1}}$, due to the dramatic decrease in particle density with height. 

The contribution function of the Fe \textsc{xii} 195.1 \mbox{\AA} line is shown in Figure~\ref{fig6} h). Compared with the emission from the coronal hole, the streamer makes a large contribution to the Fe \textsc{xii} radiation below 1.2 $R_\odot$ in the AWSoM simulation because of the higher electron density and an electron temperature that is closer to the maximum abundance temperature of Fe \textsc{xii}.

\section{Discussion}\label{sec4}
We compared the Fe \textsc{xii} 192.4, 193.5, and 195.1 \mbox{\AA} and Fe \textsc{xiii} 202.0 \mbox{\AA} line widths in the southern coronal hole observed by Hinode/EIS with the AWSoM simulations. There is no trend for the measured line widths to decrease above 1.2 $R_\odot$, which is found in some previous researches \citep[e.g.,][]{Bemporad2012,Hahn2012} with a large uncertainty on measured line widths. The measured line widths are within the uncertainty of the undamped profiles, which means that the waves might or might not be damped below 1.5 $R_\odot$. In addition, there is a larger discrepancy between the EIS observations and AWSoM simulations. 

Here we discuss a few factors that may cause the discrepancy between our observations and the AWSoM simulations or previous measurements.
\begin{figure*}[t]
	\centering
	\includegraphics[width=\linewidth]{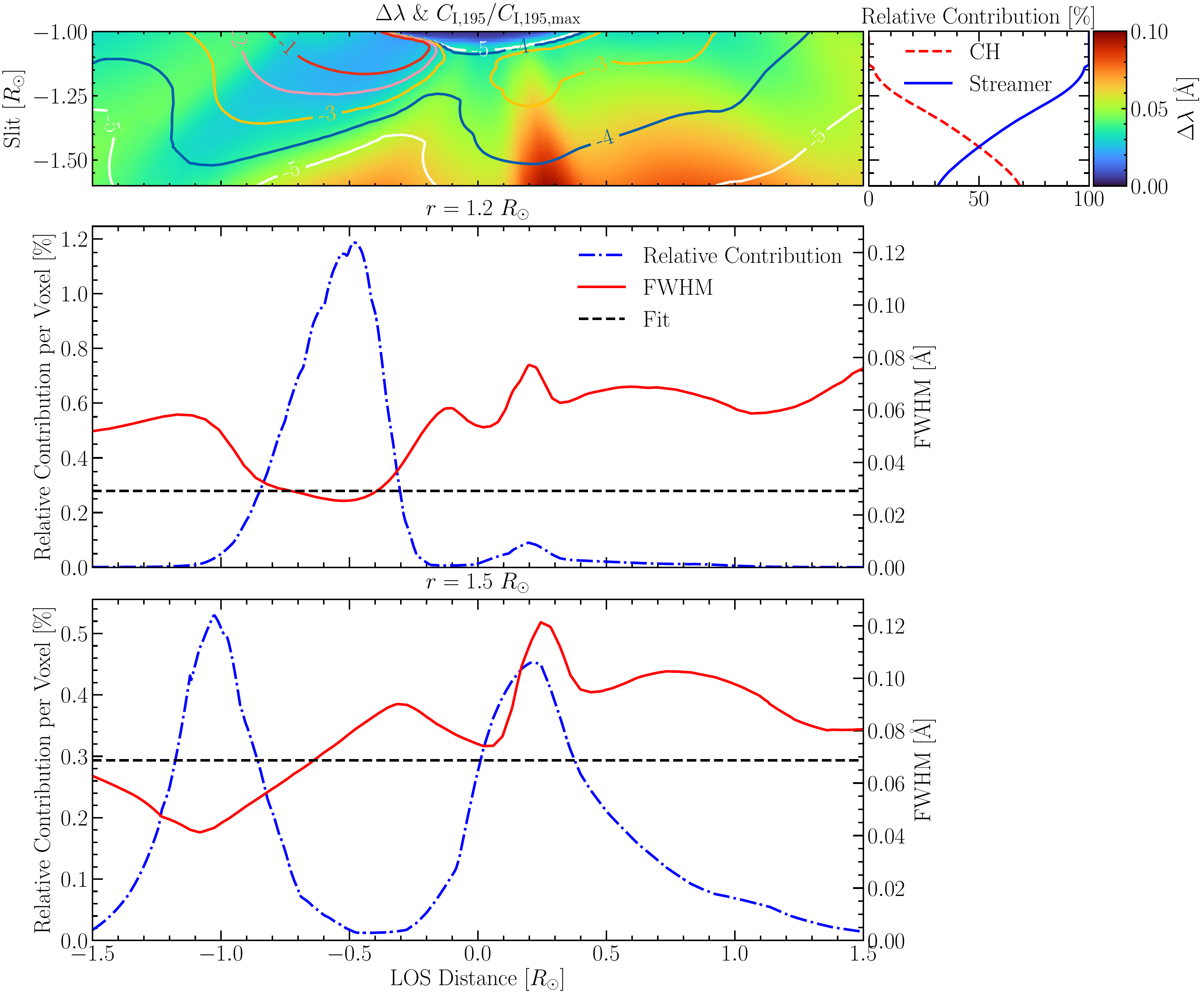}
	\caption{(Upper left) Local total FWHM $\Delta \lambda$ distribution in the meridional cut of the AWSoM results. The contours show the levels of the Fe \textsc{xii} 195.1 \mbox{\AA} contribution function normalized to its maximum $C_{\mathrm{I,195}}/C_{\mathrm{I,195,max}}$. The contour labels are in the logarithmic scale, i.e., $-1$ means $10^{-1}$ of the maximum value. (Upper right) Relative contribution of Fe \textsc{xii} 195.1 emission originated from the coronal hole or the streamer at different heights in AWSoM simulations. (Middle) Relative contribution of Fe \textsc{xii} 195.1 emission per voxel along the LOS (dot-dashed blue) and local FWHM distribution along the LOS (solid red) at a heliocentric distance of 1.2 $R_\odot$. The dashed horizontal line indicates the fitted emergent Fe \textsc{xii} line width at 1.2 $R_\odot$. (Bottom) The same as the middle panel, but at a heliocentric distance of 1.5 $R_\odot$.}
	\label{fig7}
\end{figure*}

\subsection{Streamer Contamination}

The presence of a streamer at the far side of the Sun is demonstrated in the meridional cut of the AWSoM simulations in Figure~\ref{fig6}. To understand how the streamer affects the formation of the monotonically increasing line widths in simulations, we plot the contours of the Fe \textsc{xii} 195.1 \mbox{\AA} contribution function normalized to its maximum $C_{\mathrm{I,195}}/C_{\mathrm{I,195,max}}$ over the local FWHMs $\Delta \lambda$ along the LOS in the upper left panel of the Figure~\ref{fig7}. The relative contribution of the Fe \textsc{xii} 195.1 emission from the polar coronal hole or the streamer is shown in the upper right panel. First, the local FWHMs are dominated by nonthermal velocity and increase from $\sim$0.06 \mbox{\AA} to $\sim$0.1 \mbox{\AA} in the coronal hole. Second, photons from the streamer dominate the synthetic profiles with narrower line widths, even up to 1.4 $R_\odot$. 

The middle and lower panel of Figure~\ref{fig7} show the LOS distribution of the relative contribution of the Fe \textsc{xii} 195.1 emission per voxel along the LOS (i.e., the local contribution function normalized to the total Fe \textsc{xii} 195.1 \mbox{\AA} emergent intensity) and the local FWHM at a heliocentric distance of 1.2 $R_\odot$ and 1.5 $R_\odot$. The fitted line widths from the synthetic profiles are also provided as references. The line profiles in the AWSoM simulation are much narrower in the lower corona, e.g., at 1.2 $R_\odot$, because the SPECTRUM LOS integration causes the streamer emission to provide the bulk of the observed photons. The AWSoM line widths begin to increase with height because the streamer contributes fewer photons in higher altitudes, e.g., 1.5 $R_\odot$ and the coronal hole FWHMs start to increase with height. So the fitted line width at 1.5 $R_\odot$ is $\sim$0.07 \mbox{\AA}, which is much larger than that of $\sim$0.03 \mbox{\AA} at 1.2 $R_\odot$.

The AWSoM simulation shows how the streamer photons are likely to contaminate the spectrum and result in much narrower line profiles. The simulation also shows how complex the variation of local nonthermal velocity along the LOS is even within coronal holes, which means that a simple Gaussian fitting may not explain the observed profiles. In addition, AWSoM has a limited spatial resolution and uses a synoptic magnetogram to calculate the inner boundary, which cannot resolve the small-scale effects or structures of the sizes of the FWHM fluctuations. Therefore we cannot use AWSoM to investigate the nature of the FWHM periodic spatial fluctuations in the line widths shown in Figure~\ref{fig5}.  

\begin{figure}[htb!]
	\centering
	\includegraphics[width=\linewidth]{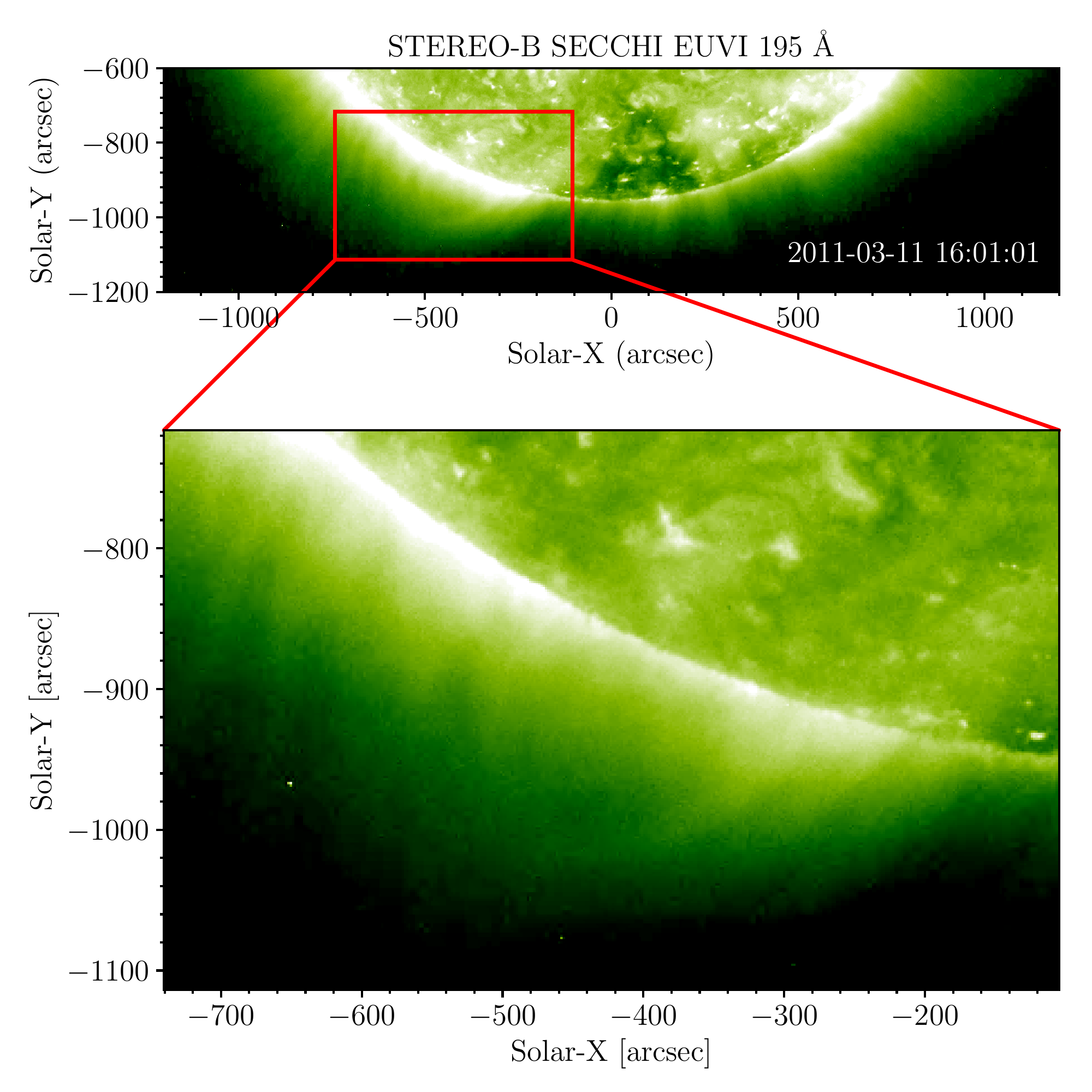}
	\caption{SECCHI EUVI 195 \mbox{\AA} imaging of the streamer on 2011 March 11, when the STEREO-B spacecraft was in the quadrature with Earth and Sun. The red rectangle in the upper panel outlines the FOV of the lower panel. Note that the intensity in the lower panel has been rescaled.}
	\label{fig8}
\end{figure}

Based on the AWSoM simulations, we suggest that the Fe \textsc{xii} and Fe \textsc{xiii} emission in the observation is also significantly contaminated by the streamer emission. In addition, the streamer was also observed by SECCHI EUVI \citep{Howard2008} 195 \mbox{\AA} imaging on board the STEREO-B spacecraft \citep{Kaiser2008}, as both the STEREO spacecrafts were in quadrature with Earth and Sun (see Figure~\ref{fig8}). The fluctuations in the line widths may come from the streamer \cite[also see][Figure 4]{Singh2003}.
 
Synthetic line widths are much narrower than the observed ones either because AWSoM underestimates the nonthermal broadening in the streamer or because the observed Fe \textsc{xii} line profiles are less contaminated by the streamer emission than predicted. The possible differences between the variation of FHWM versus height in this study and previous studies \citep[e.g.,][]{Bemporad2012,Hahn2012} may be due to (1) contamination from the streamer, (2) different wave damping levels in other coronal holes, and (3) wave-damping levels affected by different phases of the solar cycle.

\subsection{Ion Temperature in AWSoM}\label{sec4:iontemp}
The SPECTRUM module uses the LOS component of the anisotropic proton temperature to evaluate the thermal broadening of the spectral lines in place of the temperature of each ion because the current version of AWSoM does not calculate the temperature of each ion. However, a previous study \citep[e.g.,][]{Moran2003} suggested that there is no uniform ion temperature in the off-limb corona. The ion temperature could also deviate greatly from the local proton and electron temperatures due to some other heating mechanisms such as ion-cyclotron resonance \citep[e.g.,][]{Tu1998}. \citet{Landi2009} measured the Fe \textsc{xii} temperature in the coronal hole between 1.03 and 1.17 $R_\odot$ and obtained a result of $\log T_i \sim 6.7-6.95$, which corresponds to a thermal FHWM of $\sim 0.05$ \mbox{\AA}. The Fe \textsc{xii} ion temperature in the quiet solar corona is about $\log T = $6.2--6.6 \citep{Landi2007}, which is also higher than the proton temperature used in AWSoM/SPECTRUM simulations. 

\begin{figure}[]
	\centering
	\includegraphics[width=\linewidth]{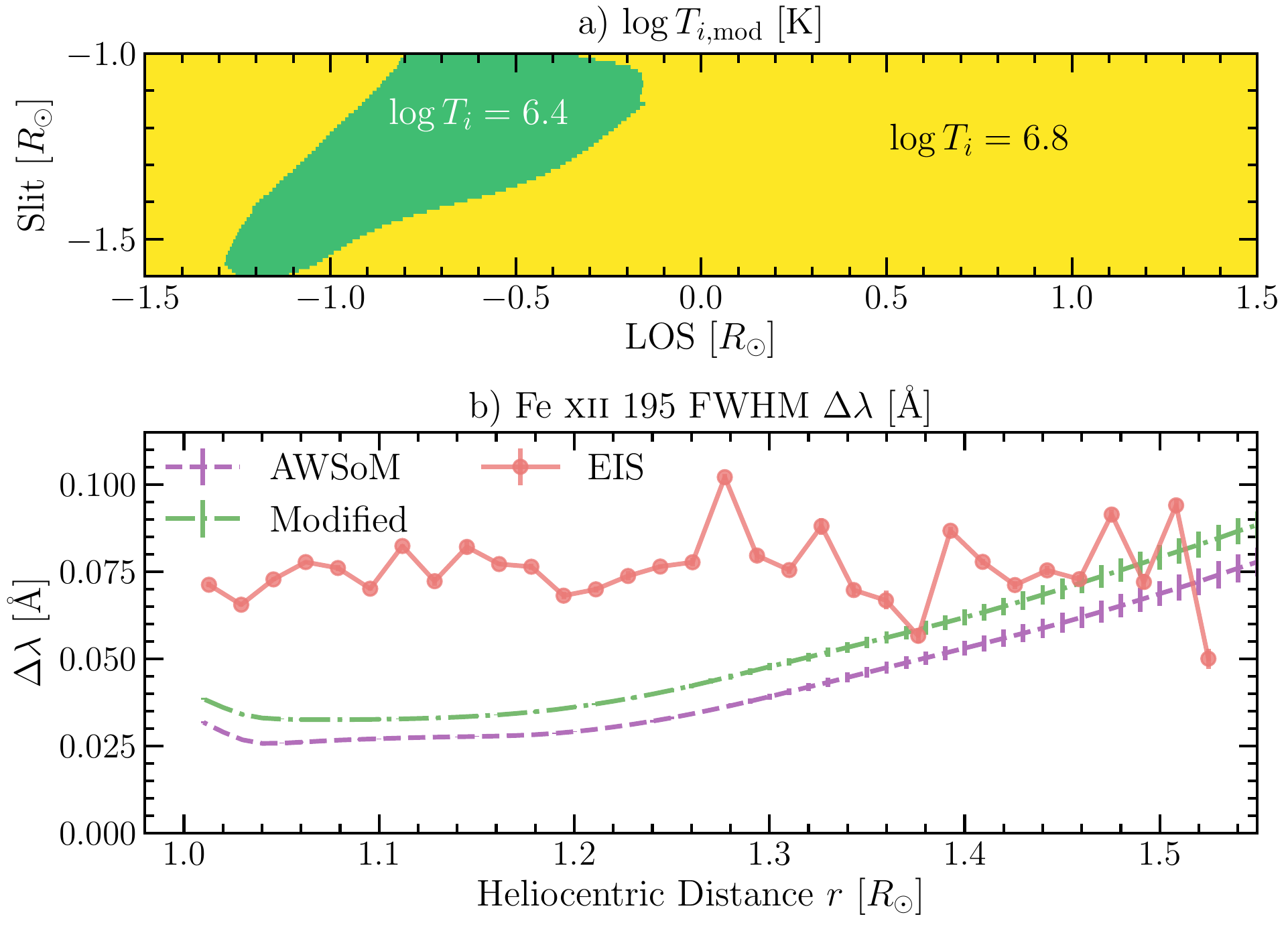}
	\caption{(a) Modified Fe \textsc{xii} ion temperature distribution in the meridional cut of the solar corona taken in the plane corresponding to the LOS, where the green region stands for the streamer and the yellow region is treated as the coronal hole. (b) Fe \textsc{xii} 195 \mbox{\AA} line widths synthesized from the modified Fe \textsc{xii} ion temperature compared with the current AWSoM simulation and EIS observations.}
	\label{fig9}
\end{figure}

To investigate the influence of a higher ion temperature on the line broadening, we manually changed the temperatures in the meridional cut, as shown in Figure~\ref{fig9} a). We first determined the streamer region using the contribution function of the Fe \textsc{xii} 195 \mbox{\AA} line and then arbitrarily assigned a temperature of $\log T = 6.4$ to this region. The ion temperature in the remaining grid is set to be $\log T = 6.8$. Then we resynthesize the Fe \textsc{xii} 195 \mbox{\AA} line profiles using the new thermal broadening. We compare the line widths measured from the modified line profiles with those from the current AWSoM simulation and EIS observations in Figure~\ref{fig9} (b). Because most of the emission in the lower corona comes from the streamer, the line widths only increase by $\sim$0.01 \mbox{\AA} below 1.35 $R_\odot$, which is still insufficient to explain the line broadening at lower altitudes. Above 1.4 $R_\odot$, the increased ion temperatures ($\log T \sim 6.8$) broaden the line profiles by $\sim$0.02 \mbox{\AA}, which causes the synthetic widths to become much closer to the EIS observations. We have to stress that these results come from a very crude and arbitrary approximation of the real ion temperatures. Nevertheless, they point toward an important parameter that could be responsible for the disagreement between the measured and observed FHWM values. Still, while the much higher ion temperature may account for the widths at heights where the streamer is not present, it cannot reproduce the widths where the streamer dominates the emission.  

\subsection{Spatial Binning and FOV} \label{sec4:spabin}
\begin{figure*}
	\centering
	\includegraphics[width=0.8\linewidth]{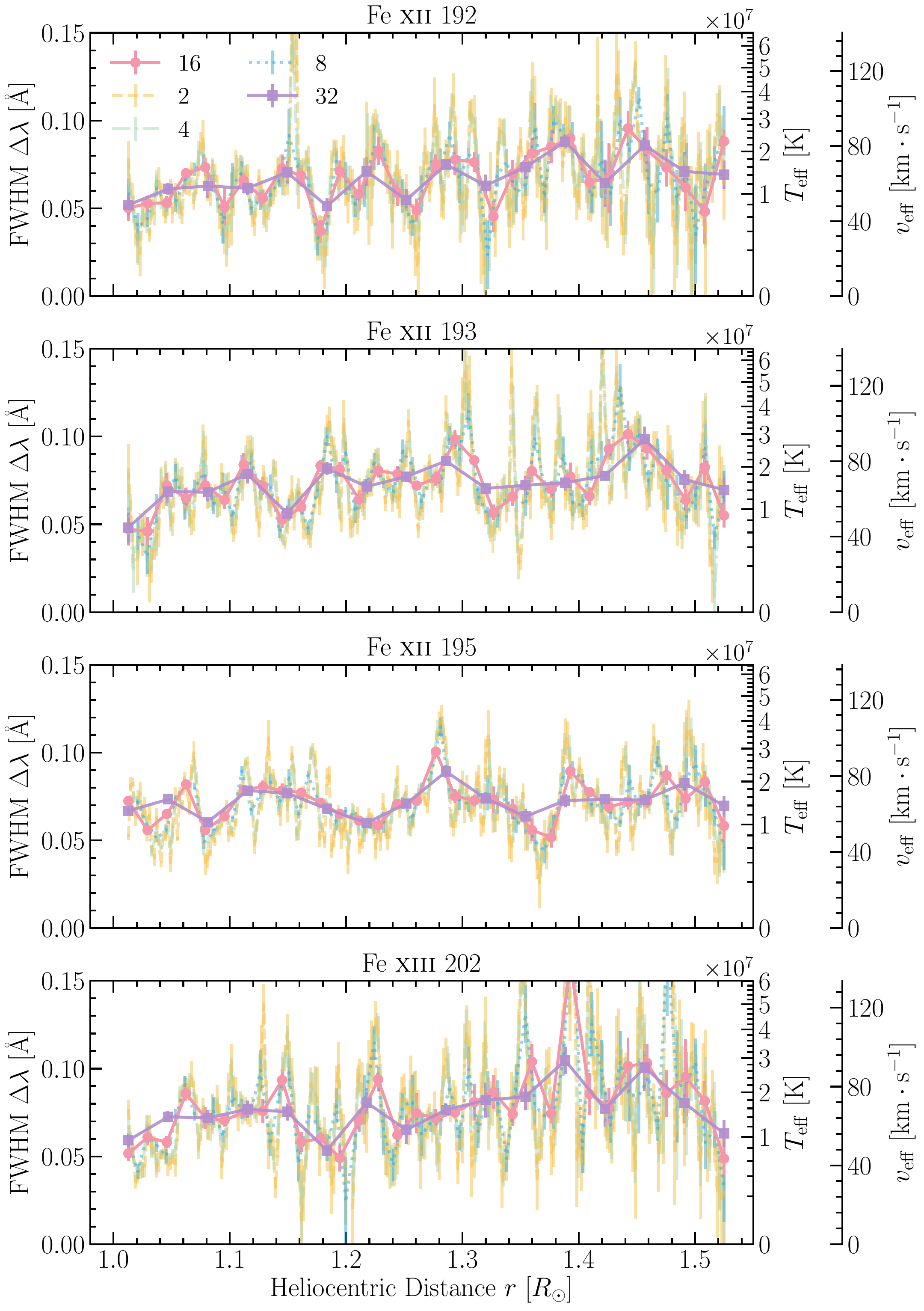}
	\caption{FWHMs of Fe \textsc{xii} 192.4, 193.5, and 195.1 \mbox{\AA} and Fe \textsc{xiii} 202.0 \mbox{\AA} measured by different spatial binnings: every 2, 4, 8, 16, and 32 pixels along the slit using a 2\% stray-light level. In order to save the computation time, the line profiles used in this figure are fit using a maximum likelihood optimization method. The error bars in this figure are from the diagonal components of the covariance metrics. }
	\label{fig10}
\end{figure*}

Previous studies have used large spatial binning, usually more than 30 pixels, to increase the S/N and obtain Gaussian profiles to fit. Because our observations have an extremely long exposure time of 33,600 s, we fitted the line profiles with different spatial binnings every 2, 4, 8, 16, and 32 pixels. The results are shown in Figure~\ref{fig10}. Smaller spatial binnings like 2 or 4 pixels provide a larger uncertainty as well as more fluctuating FHWMs. An 8- or 16-pixel binning removes the smallest-scale fluctuations due to a strong noise reduction, but it maintains variations at the 0.05--0.1 $R_\odot$ level, making them significant. However, the 32-pixel binning is so large that it smooths these fluctuating fine structures without improving the uncertainty. Therefore we suggest that the fluctuations may not be found in previous studies due to the large spatial binning, even if the emission is contaminated by other structures. 

\subsection{Stray-light Level}
\begin{figure}[]
	\centering
	\includegraphics[width=\linewidth]{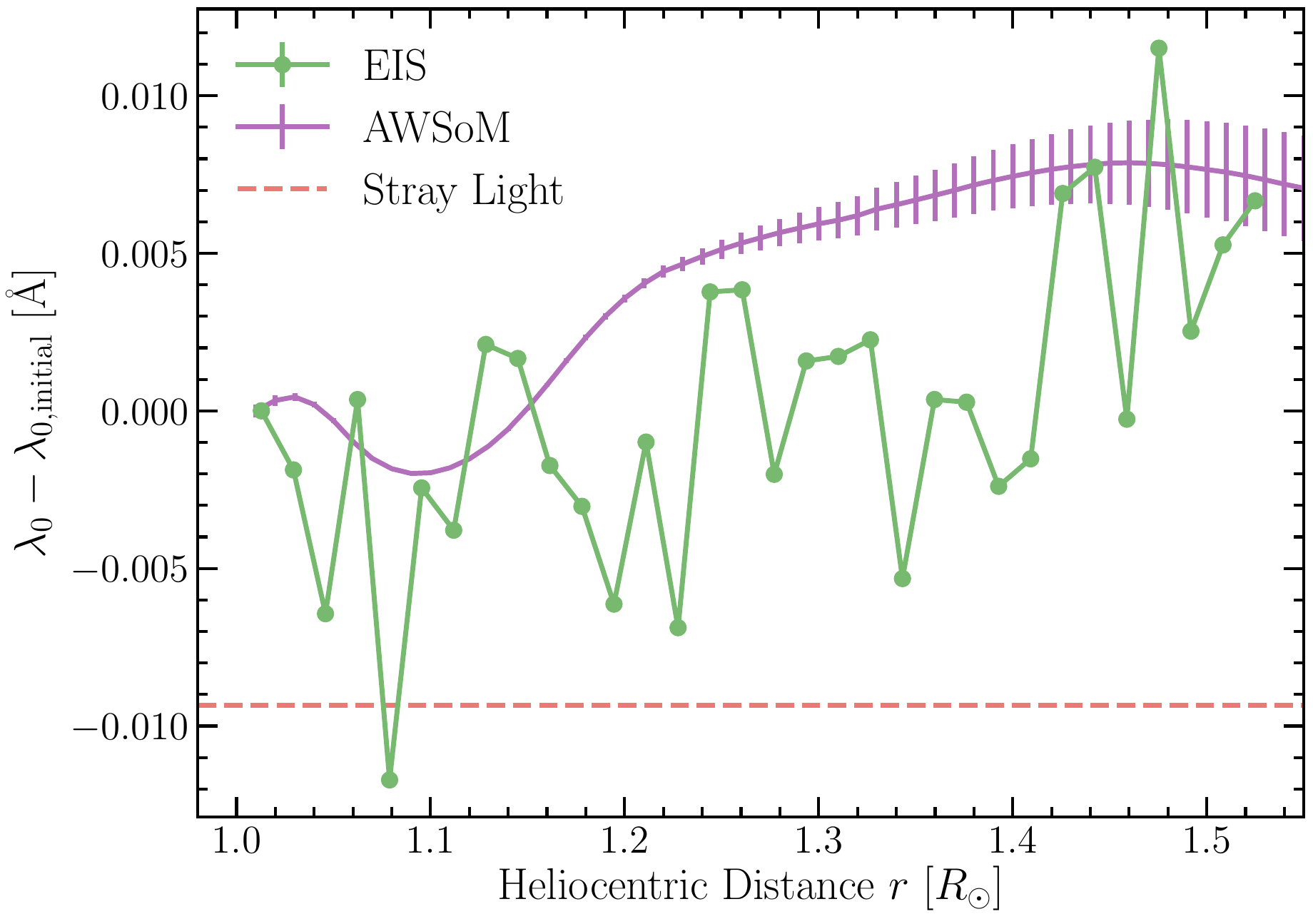}
	\caption{The variation in the Fe \textsc{xii} 195 \mbox{\AA} line centroid wavelengths $\lambda_0$ as a function of height measured from EIS observations and AWSoM simulations. We show the differences between the line centroid wavelengths $\lambda_0$ and the line centroid wavelength $\lambda_{0,\mathrm{initial}}$ of the pixel closest to the limb. The horizontal line indicates the line centroid wavelength of the fitted stray-light profile.}
	\label{fig11}
\end{figure}

The measured line centroid wavelengths from EIS observations and AWSoM simulations are shown in Figure~\ref{fig11}. By choosing different stray-light levels, we show that a larger stray-light component may result in a larger FWHM. Most of the previous studies used the 2\% stray-light level obtained from \citet{ugarte2010}. However, the wavelength offset between the stray light might be due to the real Doppler shifts in the solar wind flows. In this case, it is possible that the stray light does not dominate the off-limb spectrum even at 1.5 $R_\odot$, because otherwise, the off-limb profiles should have the same line centroid wavelength as the stray light. The line centroids in AWSoM simulations are first blueshifted because of the bulk motion in the streamer toward the observer. The AWSoM line centroid wavelength increases by $\sim 0.01$ \mbox{\AA} at higher altitudes because the photons are increasingly emitted by the redshifted and blueshifted flows from the solar wind, and less from the blue shifted streamer. Because shifts of the AWSoM line centroid show a similar trend as EIS observations, we suggest that the stray light does not significantly affect the line profiles, even at 1.5 $R_\odot$. Moreover, the 10\% stray light would make the redshift even larger and the line widths artificially narrower at large heights, so that it is likely an overestimation. After all, the off-disk configuration of the EIS slit in our observation is much more similar to that of the eclipse configuration from \citet{ugarte2010} than the full-disk configuration of \citet{Wendeln2018}. Therefore estimating of the stray-light fraction of 2\%-4\% may be sufficient in this study.

\subsection{Photoexcitation}
\begin{figure}[]
	\centering
	\includegraphics[width=\linewidth]{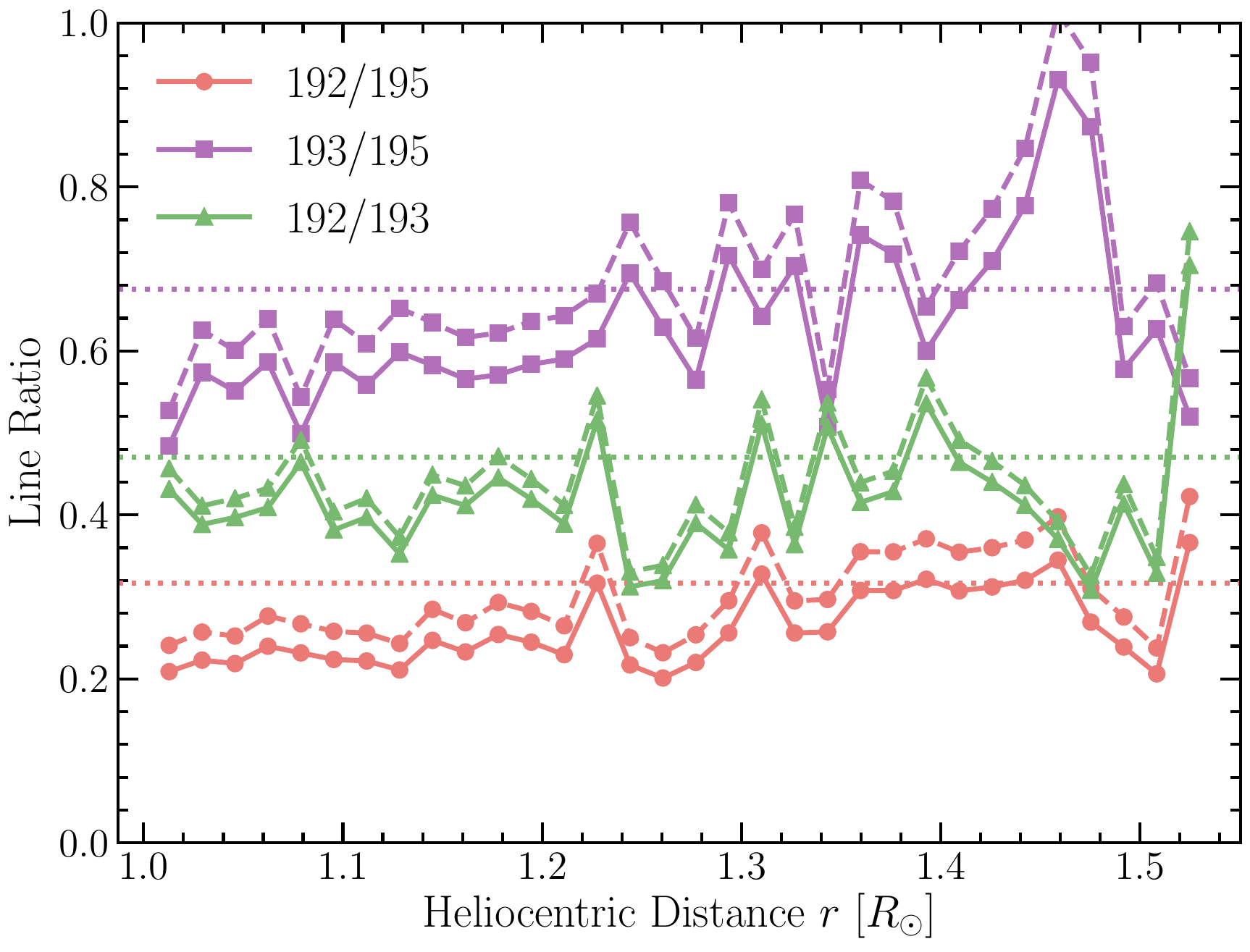}
	\caption{Fe \textsc{xii} 192.4, 193.5 and 195.1 \mbox{\AA} line intensity ratios at different height using different radiometric calibration method: \citet{Warren2014} (solid) and \citet{DelZanna2013} (dashed). The dotted horizontal lines indicate the reference values given by the CHIANTI database.}
	\label{fig12}
\end{figure}

 The Fe \textsc{xii} 192.4, 193.5, and 195.1 \mbox{\AA} line ratios may be sensitive to photoexcitation at large heights, where collisional excitation is less efficient and self-absorption of Fe \textsc{xii} emission coming from a lower altitude, brighter area could contribute to populating the parent $^4P$ levels. In this case, their ratios should be dependent on height. We show the intensity ratios of each two of the three lines in Figure~\ref{fig12} as a function of height using two different radiometric calibration methods: those of \citet{DelZanna2013} and of \citet{Warren2014}. The reference values given by CHIANTI are also plotted. The line ratios do not vary significantly below 1.2 $R_\odot$. The ratio of the Fe \textsc{xii} 192.4 and 195.1 \mbox{\AA} lines and the ratio of the Fe \textsc{xii} 193.5 and 195.1 \mbox{\AA} lines increase with height, while the Fe \textsc{xii} 192.4 and 193.5 \mbox{\AA} line ratios do not show significant variations. \citet{Hahn2012} calculated the line ratios, but found no notable changes at different heights. Because these ratios are independent of temperature and density, different structures along the LOS should not alter the ratios. Moreover, if photoexcitation were active, the brightest line (i.e., Fe \textsc{xii} 195 \mbox{\AA}) should become brighter with the height relative to the others, so that the ratios should decrease rather than increase. The systematic increase in the Fe \textsc{xii} 192/195 and 193/195 line ratios may not result from photoexcitation, but may be caused by other instrumental and physical effects.

\subsection{Instrumental Effect}\label{sec4:inseff}
\begin{figure}[]
	\centering
	\includegraphics[width=\linewidth]{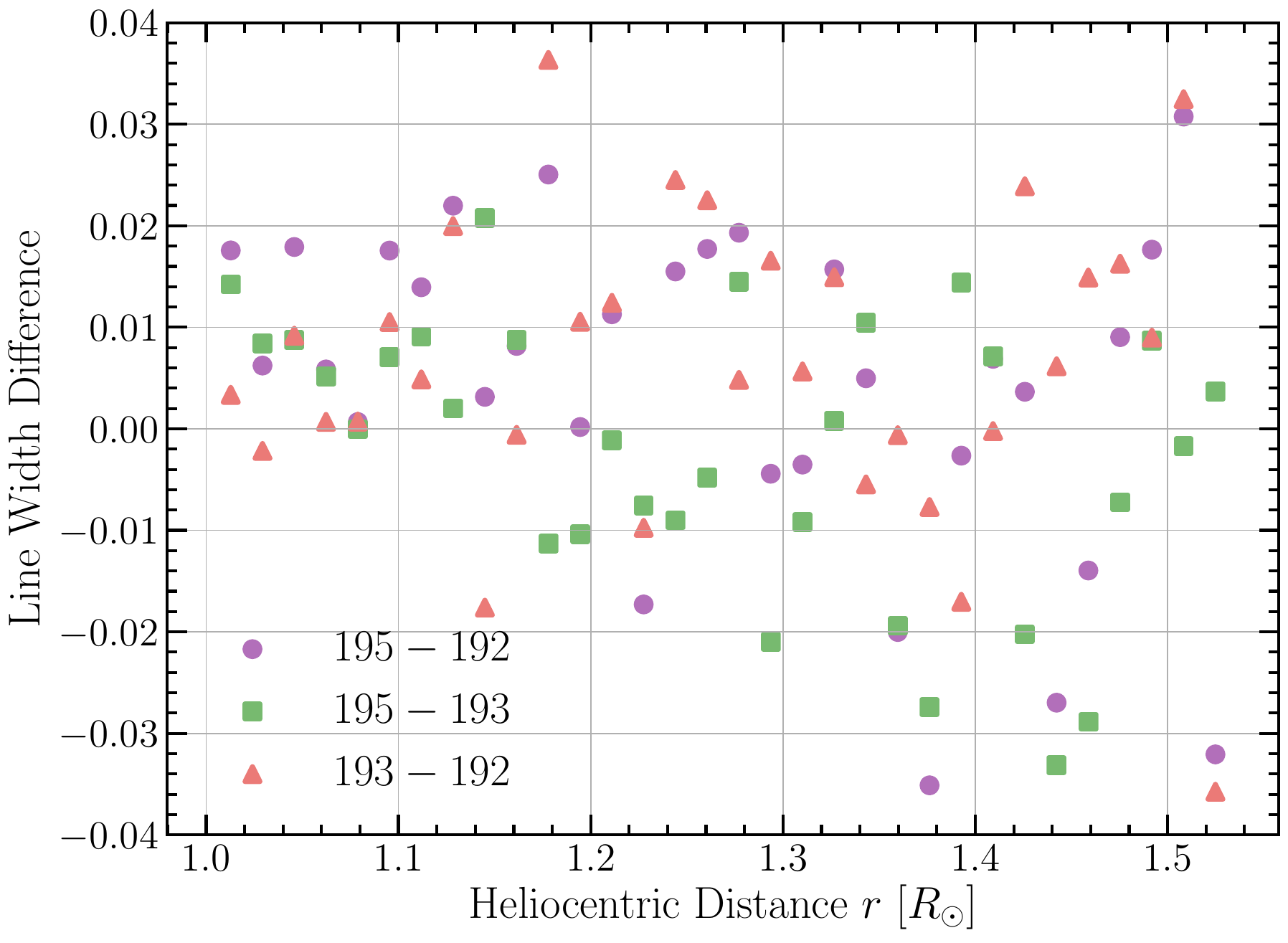}
	\caption{The difference between the FWHM of the Fe \textsc{xii} 192.4, 193.5, and 195.1 \mbox{\AA} lines at different heights. Below 1.3 $R_\odot$, most of the circle ($195-192$) and square ($195-193$) markers are above the zero.}
	\label{fig13}
\end{figure}

 Ideally, the Fe \textsc{xii} triplets should have identical line widths at any given height. However, in EIS observations, the Fe \textsc{xii} 195.12 \mbox{\AA} line is always broader than the other two. The differences in the Fe \textsc{xii} triplet line widths along the slit measured in this study are shown in Figure~\ref{fig13}. We note that below $1.2\ R_\odot$ (corresponding to CCD pixel $\sim$300-500), the Fe \textsc{xii} 195.1 \mbox{\AA} line is broader than the other two lines.
 
 \citet{Hara2019} suggested that another weaker Fe \textsc{xii} line at 195.18 \mbox{\AA} may blend with the Fe \textsc{xii} 195.12 \mbox{\AA} line and broaden the profile. The Fe \textsc{xii} 195.18 \mbox{\AA} line was found in laboratory spectra \citep{Arthanayak2020}, although some previous experiments did not resolve the line \citep[e.g.,][]{Trabert2014}.  We used CHIANTI to calculate the intensity ratio of these two lines and found that the intensity of the suggested blended line is about 2\% of the Fe \textsc{xii} 195.12 \mbox{\AA} intensity in the streamer condition ($\log n_e\sim 8.5$, measured from the Fe \textsc{xii} 195.12 to 186.86 ratio). \citet{Young2009} measured the ratio of the two blended lines by double-Gaussian fitting and showed that the Fe \textsc{xii} 195.18 to 195.12 ratio is about 5\% when $\log n_e< 9$. We also attempted to fit the Fe \textsc{xii} 195.12 \mbox{\AA} line double-Gaussian profiles, but we found the fitted Fe \textsc{xii} 195.12 \mbox{\AA} width only narrows by 0.007 \mbox{\AA}, which is insufficient to explain the maximum discrepancy by $\sim 0.02$ \mbox{\AA} between the widths of Fe \textsc{xii} 195.12 and 192.39. Furthermore, the fitted Fe \textsc{xii} 195.18 intensity is more than 10\% of the Fe \textsc{xii} 195.12 intensity, which disagrees with the density measurements. The inconsistency may be caused by the LOS integration because \citet{Young2009} used on-disk observations of an active region, or some other factors.
 
In a recent study of EIS line widths in the quiet solar corona, \citet{DelZanna2019} suggested that the anomalous widths of the strongest Fe \textsc{xii} 193.5 \mbox{\AA} and 195.1 \mbox{\AA} line are due to instrumental reasons. They called into question whether firm conclusions could be obtained because of the uncertainties on the instrumental broadening described in \citet{EISNote7}. Our results show similar patterns in the difference of the FWHMs of Fe \textsc{xii} measured by \citet[][see Figure 2]{DelZanna2019} in 2006, but with larger standard deviations. Therefore we cannot exclude the existence of the instrumental broadening that depends not only on the position along the slit, but also on wavelength.

\section{Conclusions}
We measured the Fe \textsc{xii} 192.4, 193.5, and 195.1 \mbox{\AA} and Fe \textsc{xiii} 202.0 \mbox{\AA} line widths in a southern coronal hole up to 1.5 $R_\odot$ observed by Hinode/EIS on 2011 March 11, with a total exposure time exceeding 30,000 s. The line widths in our observation first increase between 1.0--1.05 $R_\odot$ and then start to fluctuate in the $\sim 0.05$--$0.1$ \mbox{\AA} range. We found no evidence of a systematic line width decrease with height above 1.2 $R_\odot$ in the coronal hole. We compared the observations with the predictions made by the AWSoM and SPECTRUM modules. The synthetic line widths are much narrower (by $\sim 0.03$ \mbox{\AA}) than what we observed below 1.3 $R_\odot$ and increase monotonically from $\sim$0.03 to $0.07$ \mbox{\AA} between 1.0 and 1.5 $R_\odot$. We discussed the factors that may affect the line broadening. We suggested that the off-limb Fe \textsc{xii} emission in both the simulations and observations are significantly contaminated by emission from a stream along the LOS. The discrepancy between the model and observations occurs either because the AWSoM model underestimates the nonthermal broadening in the streamer or because the observations are less affected by the streamer than predicted. The AWSoM model allows us to identify the source of the EUV emissions along the LOS, which is important for off-limb observations.


\section*{ACKNOWLEDGEMENTS}
{The authors acknowledge support by the University of Michigan (Y.Zhu), NASA grant 80NSSC18K0645 (J. Szente), and NSF grants AGS 1408789, 1460170, and NASA grants NNX16AH01G, NNX17AD37G and 80NSSC18K0645 (E. Landi).

Hinode is a Japanese mission developed and launched by ISAS/JAXA, collaborating with NAOJ as a domestic partner, NASA, and UKSA as international partners. Scientific operation of the Hinode mission is conducted by the Hinode science team organized at ISAS/JAXA. This team mainly consists of scientists from institutes in the partner countries. Support for the post-launch operation is provided by JAXA and NAOJ (Japan), UKSA (U.K.), NASA, ESA, and NSC (Norway).

This work uses data obtained by the Global Oscillation Network Group (GONG) program, managed by the National Solar Observatory, which is operated by AURA, Inc. under a cooperative agreement with the National Science Foundation.  The data were acquired by instruments operated by the Big Bear Solar Observatory, High Altitude Observatory, Learmonth Solar Observatory, Udaipur Solar Observatory, Instituto de Astrofisica de Canarias, and Cerro Tololo Interamerican Observatory.

CHIANTI is a collaborative project involving George Mason University, the University of Michigan (USA), and the University of Cambridge (UK).

Our team also acknowledges high-performance computing support from Pleiades, operated by NASA's Advanced Supercomputing Division.

\software{Numpy \citep{oliphant2006guide,van2011numpy}, Scipy \citep{2020SciPy-NMeth}, Astropy \citep{astropy:2013,astropy:2018}, Sunpy \citep{sunpy_community2020}, Matplotlib \citep{Hunter2007}, emcee \citep{Foreman-Mackey2013}, corner.py \citep{corner}, CHIANTI \citep{Dere1997,Dere2019}, SolarSoft (\url{https://www.lmsal.com/solarsoft}), num2tex (\url{https://github.com/AndrewChap/num2tex}).}
\bibliography{natbib}{}
\bibliographystyle{aasjournal}
\end{document}